\input epsf.tex    

\documentclass[11pt]{article}
\usepackage{psfrag}
\usepackage{graphicx}

\newcommand{\bmat}{\left(\begin{array}}
\newcommand{\emat}{\end{array}\right)}

\def\yzero{\smash{\hbox{$y\kern-4pt\raise1pt\hbox{${}^\circ$}$}}}

\def\beq{\begin{equation}}
\def\eeq{\end{equation}}
\def\beqa{\begin{eqnarray}}
\def\eeqa{\end{eqnarray}}

\def\-{\hphantom{-}}

\def\s2{\frac{1}{2}}

\def\beq{\begin{equation}}
\def\eeq{\end{equation}}
\def\beqa{\begin{eqnarray}}
\def\eeqa{\end{eqnarray}}
\def\tr{{\rm tr \,}}

\def\IF{\relax{\rm I\kern-.18em F}}
\def\II{\relax{\rm I\kern-.18em I}}

\def\cp{{\cal P}}
\def\IC{\bf C}
\def\IZ{\bf Z}
\def\IR{\bf R}

\def\IS{\bf S}
\def\IP{\bf P}
\def\IT{\bf T}

\def\z2z2{$\IC^3/(\IZ_2\times\IZ_2)$}

\def\id{{\bf 1}}

\def\lra{\longrightarrow}

\def\Dsl{\,\raise.15ex\hbox{/}\mkern-13.5mu D} 

 \def\cp#1{\relax\ifmmode {\IP\kern-2pt{}_{#1}}\else $\IP\kern-2pt{}_{#1}$\=fi}
\newcommand{\drawsquare}[2]{\hbox{%
\rule{#2pt}{#1pt}\hskip-#2pt
\rule{#1pt}{#2pt}\hskip-#1pt
\rule[#1pt]{#1pt}{#2pt}}\rule[#1pt]{#2pt}{#2pt}\hskip-#2pt
\rule{#2pt}{#1pt}}

\newcommand{\fund}{\raisebox{-.5pt}{\drawsquare{6.5}{0.4}}}
\newcommand{\antifund}{\overline{\fund}}

\begin{document}

\makeatletter \@addtoreset{equation}{section} \makeatother
\renewcommand{\theequation}{\thesection.\arabic{equation}}

\pagestyle{empty}
\vspace*{.5in}
\rightline{CERN-PH-TH/2006-040}
\rightline{IFT-UAM/CSIC-06-12}
\rightline{\tt hep-th/0603108}
\vspace{1.5cm}

\begin{center}
\LARGE{\bf Quiver Gauge Theories at Resolved and Deformed Singularities 
using Dimers} \\[10mm]

\medskip

\large{I\~naki Garc\'{\i}a-Etxebarria$^\dagger$, Fouad Saad$^\dagger$, 
Angel 
M. Uranga$^\ddagger$} 
\\
{\normalsize {\em $^\dagger$ Instituto de F\'{\i}sica Te\'orica, C-XVI \\
Universidad Aut\'onoma de Madrid \\
Cantoblanco, 28049 Madrid, Spain \\
$^\ddagger$ TH Unit, CERN, \\
CH-1211 Geneve 23, Switzerland \\
{\tt innaki.garcia@uam.es, fouad.saad@uam.es, angel.uranga@cern.ch, 
angel.uranga@uam.es} 
\\[2mm]}}

\end{center}

\smallskip

\begin{center}
\begin{minipage}[h]{14.5cm}
{\small
The gauge theory on a set of D3-branes at a toric Calabi-Yau singularity 
can be encoded in a tiling of the 2-torus denoted dimer diagram (or brane 
tiling). We use these techniques to describe the effect on the 
gauge theory of geometric operations partially smoothing the singularity 
at which D3-branes sit, namely partial resolutions and complex 
deformations. More specifically, we describe the effect of 
arbitrary partial resolutions, including those which split the original 
singularity into two separated. The gauge theory 
correspondingly splits into two sectors (associated to branes in 
either singularity) decoupled at the level of massless 
states. We also describe the effect of complex deformations, associated to 
geometric transitions triggered by the presence of fractional branes with 
confinement in their infrared. We provide tools to easily obtain the 
remaining gauge theory after such partial confinement.
}
\end{minipage}
\end{center}

\newpage                                                        

\setcounter{page}{1} \pagestyle{plain}
\renewcommand{\thefootnote}{\arabic{footnote}}
\setcounter{footnote}{0}

\section{Introduction}

The study of the $N=1$ supersymmetric gauge theory on a stack of D3-branes 
probing a Calabi-Yau threefold conical singularity is a fruitful source of 
new insights into brane dynamics \cite{Douglas:1996sw,Douglas:1997de}, 
the AdS/CFT correspondence \cite{Klebanov:1998hh}, and its extensions to 
non-conformal situations by the addition of fractional branes 
\cite{Klebanov:2000hb}.

Although present techniques do not allow this analysis for a general 
singular Calabi-Yau
\footnote{For certain simple singularities (like complex cones over del 
Pezzo surfaces) which include some non-toric cases, general methods based 
on exceptional collections have been applied, see e.g. 
\cite{Wijnholt:2002qz,Herzog:2003zc}}, several techniques have been 
successfully applied to 
the understanding of D3-branes at toric singularities:
Partial resolutions (e.g. \cite{Morrison:1998cs,Beasley:1999uz,Feng:2000mi}), 
mirror symmetry (e.g. \cite{Hanany:2001py,Feng:2002kk}), 
un-Higgsing \cite{Feng:2002fv,Hanany:2005hq}, etc. These tools have 
provided precise checks of AdS/CFT for quiver conformal field theories
\cite{Bertolini:2004xf,Benvenuti:2004dy,Benvenuti:2005ja,Franco:2005sm,Butti:2005sw,Butti:2005vn}, 
and interesting information on related non-conformal systems
\cite{Franco:2004jz,Herzog:2004tr,Franco:2005fd,Berenstein:2005xa,Franco:2005zu,Bertolini:2005di}.

A recent great improvement in the study of the system of D3-branes at a 
toric singularity has been the introduction of the so-called brane 
tilings or dimer diagrams 
\cite{Hanany:2005ve,Franco:2005rj,Hanany:2005ss,Feng:2005gw,Franco:2006gc}. 
These techniques have provided new viewpoints on the D3-brane gauge theory 
(e.g. its moduli space, its Seiberg dualities \footnote{See  
\cite{Beasley:2001zp,Feng:2001bn,Berenstein:2002fi,Herzog:2004qw}
for earlier descriptions of Seiberg dualities for D3-branes at 
singularities.}, etc). In addition they lead to interesting mathematical 
implications, like the description of the system in the large volume 
regime via exceptional collections \cite{Hanany:2006nm}, or a 
generalization of the McKay correspondence \cite{Franco:2005rj}.

One expects that, although not completely general, toric singularities 
are representative enough for the physics of D3-branes at general 
singularities.

One of the basic features of string theory at Calabi-Yau singularities is 
the existence of localized modes which can smooth out the singularity. When 
D3-branes are located at such singularities, a natural question is what is 
the gauge theory interpretation of these smoothings (in other words, how 
do the D-branes experience this smoothing). A first kind of smoothings 
corresponds to partial resolutions and are associated to Kahler 
parameters. 
It has been known for some time that these modes couple as Fayet-Iliopoulos 
terms to the D-brane gauge theory \cite{Douglas:1997de}. In concrete 
examples it has been shown that they hence trigger a partial Higgsing 
of the gauge theory reducing it to the gauge theory in the left-over 
singularity \cite{Morrison:1998cs}. The observation that minimal partial 
resolutions (those removing a triangle from the toric diagram) admit a 
simple description in terms of brane tilings \cite{Franco:2005rj}, suggests 
the existence of a simple description of general partial resolutions in 
this language. In this paper we provide such a description, in terms 
of dimer diagram concepts and directly on the gauge theory side (by 
providing the relevant Higgsing vevs associated to a partial resolution).
Our results significantly improve the understanding of partial resolutions 
in the literature, and can be used to easily analyze complicated 
resolutions which e.g. split the original singularity into two 
singularities.

A second kind of smoothing corresponds to complex deformations. These are 
extremely interesting from the gauge theory viewpoint, since they are 
related to geometric transitions triggered by fractional branes which 
experience confinement at the infrared. The prototypical situation is the 
conifold singularity with fractional branes  
\cite{Klebanov:2000hb,Vafa:2000wi}, but similar behaviour has been  
discussed in more generality (see  e.g. 
\cite{Cachazo:2001sg}, and \cite{Aganagic:2001ug,Franco:2005fd} for 
general toric singularities).

In \cite{Franco:2005fd} the gauge theory process of confinement of 
a subset of gauge factors was translated in an ad hoc manner to the dimer 
diagram language. In this paper we provide a detailed description of the 
effect of these geometric transitions in the gauge theory, allowing us to 
derive simple dimer diagram rules to  obtain the remaining theory after 
infrared confinement of the fractional brane gauge groups. Moreover, this 
description provides a new insight into the nature of other kinds of 
fractional branes, which are known not to confine and trigger a complex 
deformation, but rather remove the supersymmetric groundstate 
\cite{Berenstein:2005xa,Franco:2005zu,Bertolini:2005di,Intriligator:2005aw}.

\medskip

The two kinds of smoothings are most easily described in terms of the web 
diagrams of the toric singularity 
\cite{Aharony:1997ju,Aharony:1997bh,Leung:1997tw}. Our main tool in 
finding the gauge theory counterpart of these geometric operations is the 
close relation between the dimer diagrams and the web diagram of the 
associated singularity \cite{Feng:2005gw,Hanany:2005ss}.

The paper is organized as follows. In Section \ref{review} we review 
some aspects of the dimer diagram techniques to study the gauge theory on 
D3-branes at toric singularities. This Section contains all the background 
material on dimers we need, so Sections \ref{splitting} and 
\ref{complexdeformations} can be read independently. In Section 
\ref{splitting} 
we describe the effect of a general partial resolution of the singularity
on the gauge theory. In Section \ref{dconi} and \ref{moreexamples} we work 
out several examples in detail, describing the effects of the partial 
resolution on the dimer diagram via simple operations in its 
zig-zag paths. In Section \ref{fieldtheory} we translate the dimer diagram 
rules for partial resolution to explicit vevs for the gauge theory 
bi-fundamentals, triggering the corresponding Higgs mechanism. The proof 
of their flatness is postponed to Appendix \ref{proof}. In 
Section \ref{matchings} we provide a different view on the dimer 
description of partial resolutions in terms of perfect matchings.
Finally, in Section \ref{fractional} we discuss partial resolution in the 
presence of fractional branes, and possible obstructions.
In Section \ref{complexdeformations} we describe complex deformations as 
geometric transitions triggered by fractional branes with infrared 
confining behaviour. In Section 
\ref{dconicomp} and \ref{moreexamplescomplex} we work out several 
examples in detail, describing the effect of such geometric transitions on 
the dimer diagram via simple operations on its zig-zag paths. In Section 
\ref{fieldcomp} we describe the field theory interpretation of the complex 
deformation, and describe in terms of the dimer diagrams the gauge 
theory analysis of branes probing the confining theory. In Section 
\ref{pmcomp} we provide a different view of the complex deformation in the 
dimer \footnote{We will frequently abuse language and simply call \emph{dimer} the
  complete dimer graph. Strictly speaking a dimer is what we call an
  edge.} in terms of perfect matchings. Finally in 
Section \ref{conclu} we offer some final remarks.

\section{Review of dimer diagrams}
\label{review}

In this Section we review some background material on dimer diagrams and 
their relevant to quiver gauge theories. Reviews of the mathematical 
aspects of dimers can be found in \cite{kenyon1,kenyon2}.

\subsection{Quiver gauge theories and dimer diagrams}
\label{quiver}

\begin{figure}[!htp]
\centering
\psfrag{w}{$W = - X_{21}X_{12}X_{23}X_{32} + X_{32}X_{23}X_{34}X_{43}$}
\psfrag{x}{$- X_{43}X_{34}X_{41}X_{14} + X_{14}X_{41}X_{12}X_{21}$}
\psfrag{QUIVER}{Quiver}
\psfrag{DIMER}{Dimer}
\includegraphics[scale=0.70]{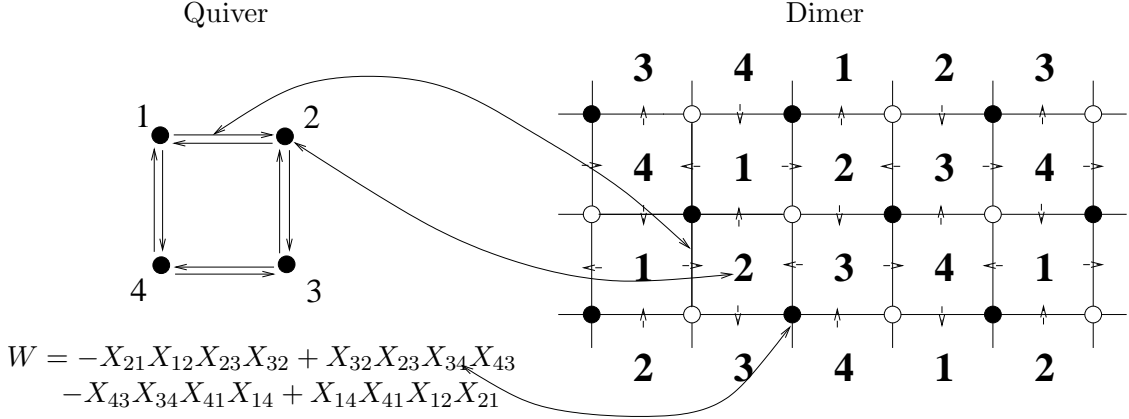}
\caption{\small Quiver and dimer for a $\IZ_2$ orbifold of the conifold. 
Faces in the 
dimer correspond to gauge groups, edges correspond to bifundamentals and 
each vertex corresponds to a superpotential term. Edges have an 
orientation determined by the coloring of the adjacent nodes.}
\label{dconidimer_quiver}
\end{figure}

The gauge theory of D3-branes probing toric threefold singularities is 
determined by a set of unitary gauge factors (of equal rank in the 
absence of fractional branes, which we do not consider for the moment), 
chiral multiplets in bi-fundamental representations, and a superpotential
given by a sum of traces of products of such bi-fundamental fields. 
The gauge group and matter content of such gauge theories can be encoded 
in a quiver diagram, such as that shown in Figure \ref{dconidimer_quiver}a,
with nodes corresponding to gauge factors, and arrows to bi-fundamentals.
The superpotential terms correspond to closed loops of arrows, but the 
quiver does not fully encode the superpotential.

Recently it has been shown that all the gauge theory information, 
including the gauge 
group, the matter content and the superpotential, can be encoded 
in a so-called brane tiling or dimer graph \cite{Hanany:2005ve,Franco:2005rj}
\footnote{The brane tiling / dimer diagram can be dualized to an improved 
quiver diagram, the periodic quiver, which also encodes all this 
information.}. This is a tiling of $\IT^2$ defined by a bipartite graph, 
namely one whose nodes can be colored black and white, with no edges 
connecting nodes of the same color \. The dictionary associates faces in the 
dimer diagram to gauge factors in the field theory, edges with 
bi-fundamental fields (fields in the adjoint in the case that the
same face is at both sides of the edge), and nodes with
superpotential terms. The bipartite 
character of the diagram is important in that it defines an orientation 
for edges (e.g. from black to white nodes), which determines the chirality 
of the bi-fundamental fields. Also, the color of a node determines the 
sign of the corresponding superpotential term.

The explicit mapping between this bipartite graph and the gauge theory, 
is illustrated in one example in Figure \ref{dconidimer_quiver}.
Many interesting features of the gauge theory have been described in terms 
of dimers by now.
%

\subsection{Dimer diagrams and the mirror Riemann surface}
\label{mirror}

There are two interesting ways to relate physically the dimer diagram with 
the gauge theory. As described in \cite{Franco:2005rj} the diagram can be 
considered to specify a configuration of NS5- and 
D5-branes, generalizing the brane box  \cite{Hanany:1997tb,Hanany:1998ru,Hanany:1998it}
and brane diamond models \cite{Aganagic:1999fe}. The NS5-branes 
extend in the $0123$ directions and wrap a holomorphic curve in the 
$4567$ directions. The D5-branes span the $012346$ directions and are 
bounded by the NS branes in the 46 directions. These directions are 
compact and parametrize a torus. The NS branes thus generate a tiling of 
this $\IT^2$, represented by a bipartite graph of the kind described 
above, hence the name brane tiling.

\begin{figure}[!htp]
\begin{center}
\psfrag{Sigma}{$\Sigma$}
\includegraphics[scale=0.5]{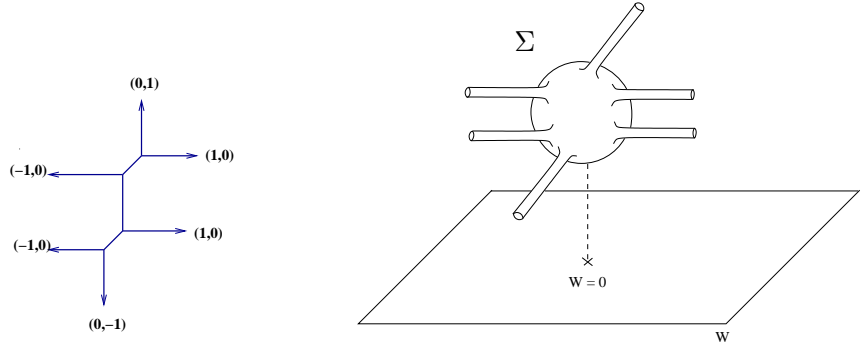}
\caption{\small a) An example of a web diagram (for the theory in Figure
\ref{dconidimer_quiver}); b) the corresponding Riemann surface $\Sigma$ 
in the mirror geometry.}
\label{thickweb}
\end{center}
\end{figure}

A second useful and more explicit viewpoint on correspondence between 
the gauge theory on D3-branes at toric singularities and dimer diagrams 
was provided in  \cite{Feng:2005gw} by using mirror symmetry, as we now 
describe \footnote{Earlier applications of mirror symmetry to D3-branes 
at singularities can be found in \cite{Hanany:2001py,Feng:2002kk}.}. The 
mirror geometry to a Calabi-Yau singularity $\mathcal{M}$ 
is specified by a double fibration over the complex plane W given by
 \beqa
 W \ &=&\ P(z,w) \ 
\\
 W \ &=& \ uv
 \eeqa
with $w,z \ \in \IC^*$ and $u,v \ \in \IC$.\
Here $P(z,w)$ is the Newton polynomial of the toric diagram of  
$\mathcal{M}$. The surface $W = P(z,w)$ describes a genus $g$ 
Riemann surface $\Sigma_W$ with punctures, fibered over $W$. The genus
$g$ equals the number of internal points of the toric diagram. The fiber 
over $W=0$, denoted simply $\Sigma$, will be important for our purposes. 
It corresponds to a smooth Riemann surface which can be thought of as a 
thickening of the web diagram 
\cite{Aharony:1997ju,Aharony:1997bh,Leung:1997tw} dual to the toric 
diagram, see Figure \ref{thickweb}.
%

\begin{figure}[!htp]
\begin{center}
\epsfxsize=8cm
\hspace*{0in}\vspace*{.2in}
\epsffile{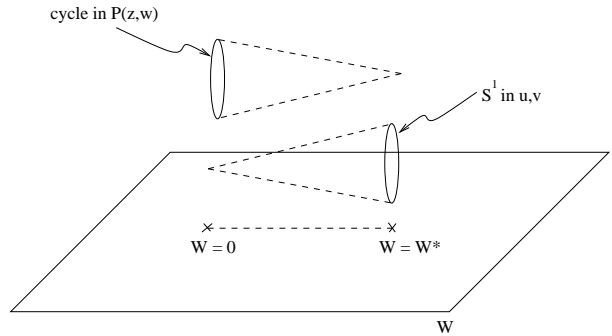}
\caption{\small Structure of the non-trivial 3-cycles in the geometry 
$\mathcal{W}$. They are constructed by fibering over the segment 
joining $W=0$ and $W=W^*$, the $S^1$ in the $uv$ fiber (degenerating at 
$W=0$) times the 1-cycle in the $P(z,w)$ fiber degenerating at $W=W^*$.}
\label{s3fiber2}
\end{center}
\end{figure}

At critical points $W=W^*$, a cycle in $\Sigma_W$ degenerates and pinches 
off. Also, at $W=0$ the $S^1$ in $W=uv$ degenerates. One can use these 
degenerations to construct non-trivial 3-cycles in the mirror geometry as 
follows. Consider the segment in the $W$-plane which joins $W=0$ with one 
of the critical points $W=W^*$, and fiber over it the $\IS^1$ in $W=uv$ 
times the 1-cycle in $\Sigma_W$ degenerating at $W=W^*$,
see Figure \ref{s3fiber2}. The result is a 3-cycle with an $\IS^3$ 
topology. The number of such degenerations of $\Sigma_W$, and hence the 
number of such 3-cycles, is given by twice the area of the toric diagram.

Mirror symmetry specifies that the different gauge factors on the 
D3-branes in the original singularity arise from D6-branes 
wrapping the different 3-cycles. The 3-cycles on which the D6-branes wrap 
intersect over $W=0$, precisely at the intersection points of the 
1-cycles in $\Sigma_{W=0}$. Open strings 
at such intersections lead to the chiral bi-fundamental fields. Moreover, 
disks in $\Sigma$ bounded by pieces of different 1-cycles lead to 
superpotential terms generated by world-sheet instantons.

Hence, the structure of the 3-cycles, and hence of the gauge theory, is 
determined by the 1-cycles in the fiber $\Sigma$ over $W=0$. This 
structure, which is naturally embedded in a $\IT^3$ (from the $\IT^3$ 
fibration structure of the mirror geometry), admits a natural projection 
to a $\IT^2$, upon which the 1-cycles end up providing a tiling of $\IT^2$ 
by a bipartite graph, which is precisely the dimer diagram of the gauge 
theory. 

\begin{figure}[!htp]
\begin{center}
\epsfxsize=10cm
\hspace*{0in}\vspace*{.2in}
\epsffile{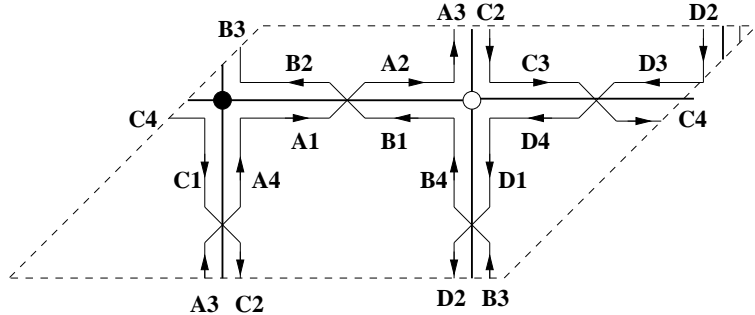}
\caption{\small Dimer of the conifold with the corresponding zig-zag paths.}
\label{conidimer_zigzag}
\end{center}
\end{figure}

This last process is perhaps better understood (and of more practical use) 
by recovering the Riemann surface $\Sigma$ from the dimer diagram of the 
gauge theory, as follows. Given a dimer diagram, one can define 
zig-zag paths (these, along with the related rhombi paths, were
introduced in the mathematical literature on dimers in \cite{kenyon1,kenyon2},
and applied to the quiver gauge theory context in
\cite{Hanany:2005ss}), as paths composed of edges, and which turn
maximally to the right at e.g. black nodes and maximally to the left at 
white nodes. They can be conveniently shown as oriented lines that cross 
once at each edge and turn at each vertex, as shown in Figure 
\ref{conidimer_zigzag}.
Notice that at each edge two zig-zag paths must have opposite 
orientations. For dimer models describing toric gauge theories, these 
zig-zag paths never intersect themselves and form closed loops wrapping 
$(p,q)$ cycles on the $\IT^2$. This is shown for the conifold in  
Figure \ref{conidimer_zigzag} where the zig-zag paths A, B, C and D have 
charges (0,1), (-1,1), (1,-1), (0,-1) respectively.

\begin{figure}[!htp]
\begin{center}
\epsfxsize=10cm
\hspace*{0in}\vspace*{.2in}
\epsffile{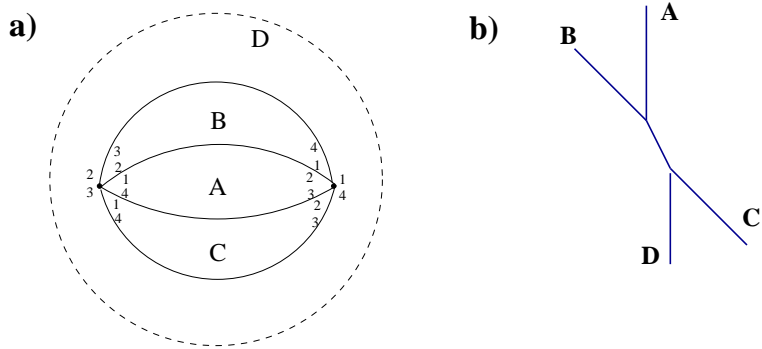}
\caption{\small a) Tiling of the Riemann surface (which is 
topologically a sphere, shown as the complex plane) for the case of 
D3-branes at a conifold singularity. b) The web diagram, providing a 
skeleton of the Riemann surface, with asymptotic legs 
corresponding to punctures (and hence to faces of the tiling of $\Sigma$, 
and zig-zag paths of the original dimer diagram).}
\label{coniriem}
\end{center}
\end{figure}

As shown in \cite{Feng:2005gw}, the zig-zag paths of the dimer diagram 
associated to D3-branes at a singularity lead to a tiling of the Riemann 
surface $\Sigma$ in the mirror geometry. Specifically, each zig-zag path 
encloses a face of the tiling of $\Sigma$ which includes a puncture, and 
the $(p,q)$ charge of the associated leg in the web diagram is the $(p,q)$ 
homology charge of the zig-zag path in the $\IT^2$. The touching of 
two of these faces in the tiling of $\Sigma$ corresponds 
to the coincidence of the corresponding zig-zag paths along an edge of the 
dimer diagram. The tiling of $\Sigma$ for the conifold is shown in Figure 
\ref{coniriem}a, while the corresponding web diagram is shown in Figure 
\ref{coniriem}b.

\begin{figure}[!htp]
\begin{center}
\epsfxsize=10cm
\hspace*{0in}\vspace*{.2in}
\epsffile{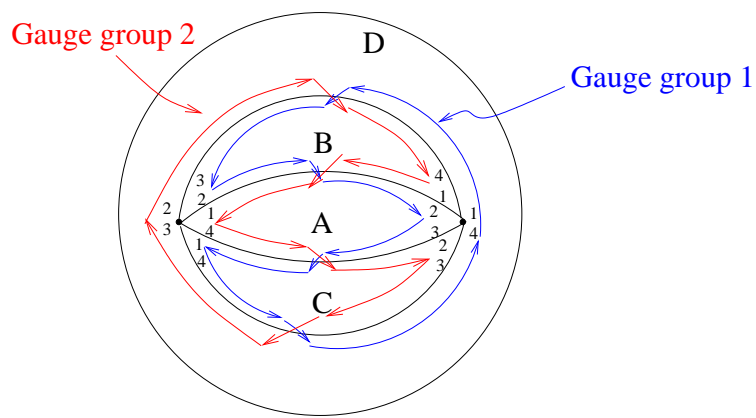}
\caption{\small Tiling of the Riemann surface for the case of D3-branes 
at a conifold singularity, with the 1-cycles corresponding to the two 
gauge factors (shown as zig-zag paths of the tiling of $\Sigma$).}
\label{conisurf}
\end{center}
\end{figure}

The dimer diagram moreover encodes the 1-cycles in the mirror Riemann 
surface, associated to the different gauge factors in the gauge theory. 
Consider a gauge factor associated to a face in the dimer diagram. One can 
consider the ordered sequence of zig-zag path pieces that appear on the 
interior side of the edges enclosing this face. By following these pieces 
in the tiling of $\Sigma$ one obtains a non-trivial 1-cycle in $\Sigma$ 
which corresponds precisely to that used to define the 3-cycle wrapped 
by the mirror D6-branes carrying that gauge factor. Using this map, it is 
possible to verify all dimer diagram rules (edges are bi-fundamentals, 
nodes are superpotential terms) mentioned at the beginning. An amusing 
feature is that these non-trivial 1-cycles in $\Sigma$ are given by
zig-zag paths of the tiling of $\Sigma$. The non-trivial 1-cycles in the 
mirror Riemann surface for the case of the conifold are shown in Figure
\ref{conisurf}.

\subsection{Perfect matchings}
\label{pmatchings}

\begin{figure}[!htp]
\begin{center}
\epsfxsize=6cm
\hspace*{0in}\vspace*{.2in}
\epsffile{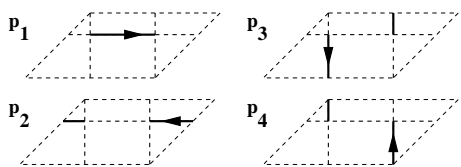}
\caption{\small Perfect matchings for the dimer of the conifold.}
\label{conipmatch}
\end{center}
\end{figure}

A last concept we would like to discuss is that of perfect matchings for a 
dimer diagram. A perfect matching is a subset of edges of the dimer 
diagram, such that every vertex of the graph is the endpoint of exactly 
one such edge. In Figure \ref{conipmatch} we show the four perfect 
matchings for 
the conifold. For future convenience, we consider the edges in each 
perfect matching to carry an orientation, e.g. from black to white nodes.
%
%

As discussed in \cite{Franco:2005rj} and proved in \cite{Franco:2006gc},
there is a one to one correspondence 
between the perfect matchings for a dimer diagram and the linear sigma 
model fields that arise in the construction of the moduli space of the 
quiver gauge theory \cite{Douglas:1997de,Morrison:1998cs,Feng:2000mi}. 
This implies that each perfect matching has an associated location in the 
toric diagram of the corresponding singularity. This can be obtained as 
follows. Fix a given perfect matching as reference matching, denoted 
$p_0$. Then for any perfect matching $p_i$ we can consider the 
path $p_i-p_0$, obtained by superimposing the edges of $p_i$ and those of 
$p_0$, with flipped orientation for the latter. With the convention that 
repeated edges with opposite orientation annihilate, we obtain a 
(possibly trivial, or even empty) path in the dimer diagram, carrying a 
(possibly trivial) $\IT^2$ homology charge $(n_i,m_i)$. Then the location 
of the matching $p_i$ in the toric diagram is given by $(-m_i,n_i)$.
This definition is equivalent to that using the height functions, so
we denote $(h_{x,i},h_{y,i})=(-m_i,n_i)$ the slope of $p_i$.
Clearly the choice of reference matching simply amounts to a choice of 
origin in the toric diagram. 

For completeness, let us mention the direct definition of the slope for 
a perfect matching, see e.g. \cite{Franco:2005rj}. The dimer path 
$p_i-p_0$ 
divides the infinite tiling of $\IR^2$ into regions that can be labeled by 
an integer, with the rule that each crossing of the path changes the 
label by one unit (up- or downwards depending on orientation of the 
crossing). This label assignment, when regarded in the torus, is 
multivalued. The holonomies around the two fundamental 1-cycles are 
denoted $(h_x,h_y)$ and are called the slope of $p_i$.

\begin{figure}[!htp]
\begin{center}
\epsfxsize=13cm
\hspace*{0in}\vspace*{.2in}
\epsffile{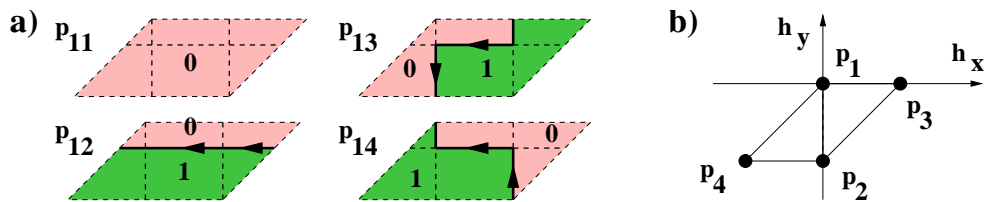}
\caption{\small 
The paths $p_{1i}=p_i-p_1$ for the dimer of the conifold are
associated to  specific locations in the toric diagram, as determined by 
the slopes, or 
equivalently by the integers $(-m,n)$ where $(n,m)$ are the homology 
charges of $p_i-p_1$.}
\label{coniheight}
\end{center}
\end{figure}

In Figure \ref{coniheight}a we have shown the 
paths $p_i-p_1$ in the dimer diagram for the conifold (along with the 
labeling by integers to obtain the slopes). The location of the perfect 
matchings in the toric diagram in shown in Figure \ref{coniheight}b.

\begin{figure}[!htp]
\begin{center}
\epsfxsize=10cm
\hspace*{0in}\vspace*{.2in}
\epsffile{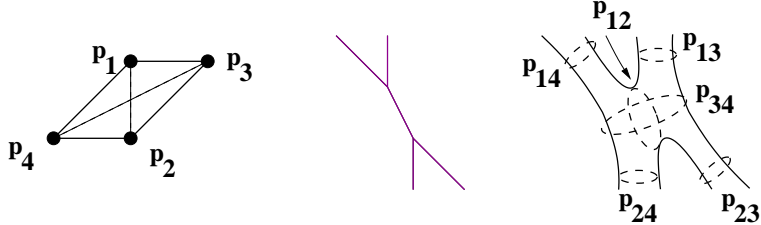}
\caption{\small The perfect matchings for the dimer of the conifold
are associated to specific locations in the toric diagram, as
determined by the slopes. The paths $p_{ij}=p_i-p_j$ correspond to 1-cycles 
in the mirror Riemann surface wrapped around the tubes dual to the segment
joining $p_i$ and $p_j$ in the toric diagram.}
\label{conipmatch2}
\end{center}
\end{figure}

Although not emphasized in the literature, there is a beautiful 
interpretation of pairs of perfect matchings. From a construction similar 
to the above, to any pair of perfect matchings $p_i$, $p_j$ one can 
associate a path (which we call `difference path') $p_{ij}=p_j-p_i$ in the 
dimer diagram, with $\IT^2$ homology 
charge $(\Delta n, \Delta m)$. In the toric diagram this is associated to 
the segment joining the location of $p_i$ to that of $p_j$, which as a 
vector is given by the slope difference
$(\Delta h_x,\Delta h_y)=(-\Delta m, \Delta n)$. Now clearly, the homology 
charge $(\Delta n, \Delta m)$ is precisely the $(p,q)$ charge of the 
segment in the web diagram dual to that segment in the toric diagram. 
This suggests a natural interpretation of $p_j-p_i$ in 
the mirror Riemann surface. Indeed, by lifting the dimer path $p_j-p_i$ to 
the mirror Riemann surface (using the tiling of the latter) one obtains a 
non-trivial 1-cycle which winds around the tube corresponding to the 
thickening of the leg in the web diagram. This is illustrated in Figure 
\ref{conipmatch2} for the case of the conifold. 

%

Clearly, the dimer paths 
associated to adjacent external matchings (i.e. matchings which are 
at adjacent locations on the boundary of the toric diagram) carry the same 
charges as zig-zag paths (although in general may not coincide edge by 
edge with them). This hence shows the equivalence of the two ways we have 
described to obtain the toric diagram associated to a dimer diagram 
\footnote{Yet another equivalent way, not used in this paper, is the 
computation of the determinant of the Kasteleyn matrix, see e.g. 
\cite{Hanany:2005ve,Franco:2005rj}.} (namely, construction of the web 
diagram by using charges of zig-zag 
paths, and construction of the toric diagram using height functions).

\section{Partial resolution}
\label{splitting}

In this Section we provide a description of the effect of general partial 
resolutions on the gauge theory using dimer diagram techniques. The 
simplest class of partial resolutions corresponds to the removal of a 
triangle in the toric diagram. There are however more involved 
possibilities, like the splitting of a singularity into two singularities 
(examples will come later). 
Clearly, the former can be regarded as a particular case of the latter, 
with the second `singularity' being a smooth patch of the final geometry.

Minimal partial resolutions (those removing one triangle of the toric 
diagram) and their relation to the D3-brane gauge theory (appearance of a 
Fayet-Iliopoulos term triggering a Higgs mechanism) were discussed in 
\cite{Morrison:1998cs}. This process was described as removal of edges in 
the dimer diagram in \cite{Franco:2005rj}. However, in both cases the 
mapping between a particular operation on the gauge theory/dimer diagram 
and the resulting geometry, was manifest only upon computation of the 
moduli space of the gauge theory. This makes it difficult to obtain the 
gauge theory corresponding to a given partial resolution, and requires 
some trial and error. Also, the description of more involved partial 
resolutions (splitting the singularities) and their gauge theory/dimer 
diagram counterpart was not provided. Moreover, as pointed out and 
explained in \cite{Hanany:2005ss}, arbitrary addition/removal of edges in 
a dimer diagram can lead to inconsistent theories.

In this Section we consider an arbitrary partial resolution of a toric 
singularity, typically splitting it into several. We consider the original
set of D3-branes to split accordingly into subsets located at the daughter 
singular points. Hence one expects that the original gauge theory splits 
(via a Higgs mechanism) into several gauge sectors, decoupled at the level 
of massless modes, and correspondingly that the original dimer diagram 
splits into several sub-dimers associated to the subsets of 
D3-branes at the daughter singularities. We provide a simple 
construction of the splitting of dimer diagrams that corresponds to a 
given partial resolution. In addition, we provide a simple recipe for the 
bi-fundamental vevs that trigger the corresponding Higgsing in the gauge 
theory. 

As a prototypical example we consider partial resolutions 
splitting a singularity into two. Other cases, like minimal partial 
resolutions, can be recovered as a particular case as mentioned 
above. Splitting into more than two daughter singularities can be easily 
obtained by iteration of our procedure.

\subsection{An example in detail}
\label{dconi}

Let us start with a simple example of the splitting via partial resolution 
of a singularity into two singularities, using concepts and techniques 
from dimers.

\begin{figure}
\begin{center}
\epsfxsize=6cm
\hspace*{0in}\vspace*{.2in}
\epsffile{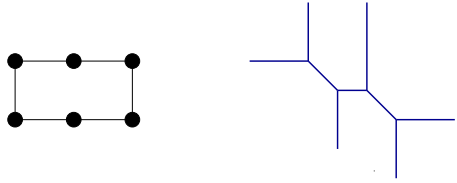}
\caption{\small The toric diagram and web diagram of the double
conifold singularity $xy=s^2w^2$. For clarity, we show the web 
diagram for a slightly resolved geometry.}
\label{dconitoric}
\end{center}
\end{figure}

\begin{figure}
\begin{center}
\epsfxsize=7cm
\hspace*{0in}\vspace*{.2in}
\epsffile{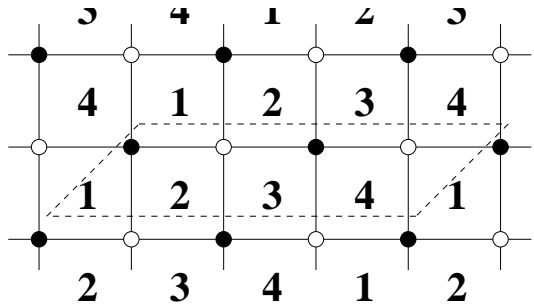}
\caption{\small Dimer diagram corresponding to the double conifold 
singularity in 
Figure \ref{dconitoric}. The dashed line corresponds to the unit cell 
of the periodic tiling.}
\label{dconidimer}
\end{center}
\end{figure}

Consider the singularity whose toric diagram and web diagram are shown in 
Figure \ref{dconitoric}. This singularity, and the gauge theory on 
D3-branes located on it, have been studied in 
\cite{Uranga:1998vf,Aganagic:1999fe}. We refer to it as the double  
conifold. The dimer diagram shown in Figure \ref{dconidimer} provides (a 
toric phase of) the gauge theory on D3-branes at this double conifold 
singularity.

\begin{figure}
\begin{center}
\epsfxsize=10cm
\hspace*{0in}\vspace*{.2in}
\epsffile{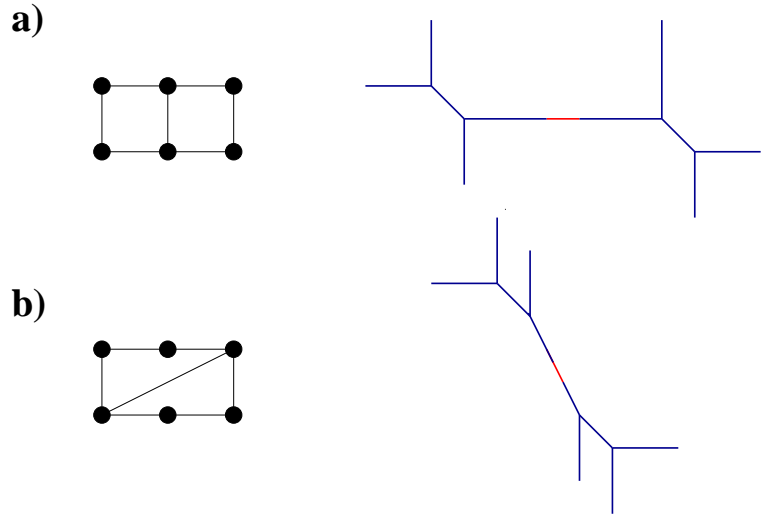}
\caption{\small Partial resolution of the double conifold singularity in 
figure \ref{dconitoric}, splitting the initial singularity into (a) 
two isolated 
conifold singularities; (b) two $\IC^2/\IZ_2$ (times $\IC$) singularities. 
The distance between the daughter singularities is controlled by the size 
of the $\IS^2$ corresponding to the dashed red segment in the associated 
web diagram. For clarity the web diagrams of the left-over singularities 
are shown for slightly resolved geometries.}
\label{dconisplit}
\end{center}
\end{figure}

The above singularity admits partial resolutions to geometries with two 
separated singularities. One such partial resolution is illustrated in 
Figure \ref{dconisplit}a, and corresponds to a large blow up of an 
$\IS^2$, smoothing the initial geometry to two isolated conifold 
singularities. A different splitting, into two $\IC^2/\IZ_2$ (times $\IC)$ 
singularities, is shown in Figure \ref{dconisplit}b.

The partial resolution corresponds to turning on Fayet-Iliopoulos terms in 
the D3-brane gauge theory \footnote{\label{uunos} This is if the $U(1)$ 
factors are included in the gauge theory. In fact, these $U(1)$ factors are 
generically massive, hence disappear from the low energy effective 
theory. In this viewpoint, the partial resolution corresponds 
to vevs for suitable baryonic operators. We however find it more 
convenient to include the $U(1)$'s explicitly, and consider the couplings 
rendering them massive at a subsequent stage. See 
\cite{Ibanez:1998qp,Morrison:1998cs} for a more detailed discussion.}. 
These FI terms force some 
of the bi-fundamental scalars to acquire a vev, breaking the gauge symmetry. 
For the case of a partial resolution splitting a singularity, the left 
over field theory must correspond to two gauge sectors, corresponding to 
the gauge theories on stacks of D3-branes at the two singularities. These 
two sectors are decoupled at the level of massless states. Namely, the 
only states charged under both sectors are massive, with mass controlled 
by the bi-fundamental vevs, and hence to the size of the 2-sphere 
responsible for the splitting. This agrees with the picture of open 
strings stretching between the two stacks of D3-branes.
In section \ref{fieldtheory} we will be more explicit about the
precise set of vevs corresponding to splitting singularities.

In this section, our aim is to provide a simple recipe that implements the 
effect of the resolution on the gauge field theory. This will be 
expressed in terms of a simple operation that, starting from the dimer 
of the initial singularity, leads to two sub-dimers corresponding to the 
gauge theories in the two daughter singularities. 

\begin{figure}
\begin{center}
\epsfxsize=12cm
\hspace*{0in}\vspace*{.2in}
\epsffile{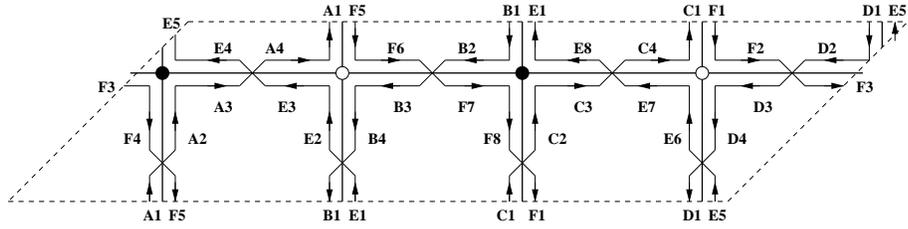}
\caption{\small Zig-zag paths for the dimer diagram of the 
double conifold.}
\label{dconipath}
\end{center}
\end{figure}

\begin{figure}
\begin{center}
\epsfxsize=10cm
\hspace*{0in}\vspace*{.2in}
\epsffile{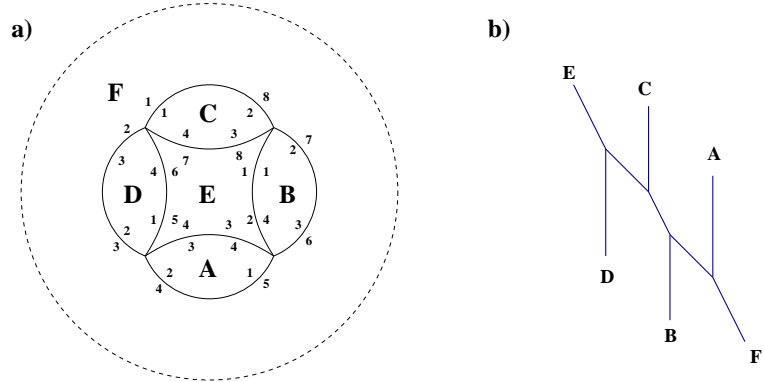}
\caption{\small (a) Zig-zag paths in Figure \ref{dconipath} 
correspond 
to external legs in the web diagram of the singularity. (b) The adjacency 
relations among zig-zag paths encode a tiling of the mirror Riemann 
surface $\Sigma$, which in this case corresponds to a 2-sphere with six 
punctures, realized in the picture as the complex plane (with the point 
at infinity).}
\label{dconiriem}
\end{center}
\end{figure}

The geometrical effect of partial resolutions is most manifest in the web 
diagram. Let us for concreteness consider the partial resolution of the 
double conifold to two conifolds, Figure \ref{dconisplit}a. As
described in Section \ref{mirror} the web diagram is encoded 
in the dimer diagram via its structure of zig-zag paths
\cite{Hanany:2005ss,Feng:2005gw}. The zig-zag paths 
corresponding to the dimer in Figure \ref{dconidimer} are shown 
in Figure \ref{dconipath}. The corresponding asymptotic legs in the web 
diagram, and the tiling of the mirror Riemann surface $\Sigma$, are shown 
in Figure \ref{dconiriem}.

\begin{figure}
\begin{center}
\epsfxsize=12cm
\hspace*{0in}\vspace*{.2in}
\epsffile{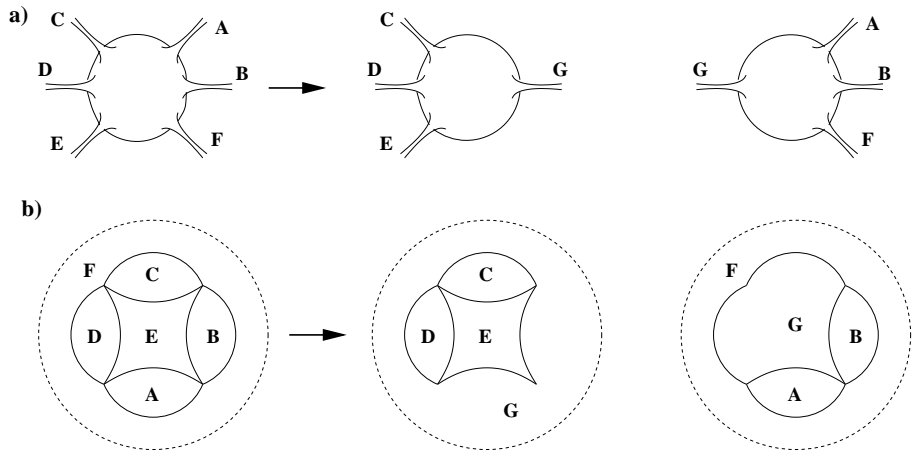}
\caption{\small (a) Schematic representation of the factorization of the 
mirror Riemann surface $\Sigma$. (b) Decomposition of the tiling of 
$\Sigma$ upon factorization. The two new sets of zig-zag paths, and their 
adjacency relations, can be used to construct the two dimers corresponding 
to D3-branes at the two singularities after splitting of the geometry.}
\label{dconifactor}
\end{center}
\end{figure}

In this language, it is easy to realize that the partial resolution 
corresponds to factorizing the Riemann surface $\Sigma$ by an elongated 
tube, as in Figure \ref{dconifactor}a. The structure of the two left over 
singularities can be determined by analyzing the local structure of the 
two daughter Riemann surfaces. Due to the factorization along the 
infinite tube, each daughter Riemann surface has a new puncture, denoted 
G, which must correspond to a new zig-zag path in the corresponding 
daughter dimer diagram. In particular, the decomposition of the 
tiling of $\Sigma$ upon this factorization, shown in Figure 
\ref{dconifactor}b, leads to two sets of zig-zag paths, namely C, D, E, G 
and A, B, F, G, respectively, with specific 
adjacency relations. This information can be used to construct two  dimer 
diagrams, which encode the gauge theories on D3-branes at the two singular 
points in the geometry after partial resolution. 

\begin{figure}
\begin{center}
\epsfxsize=12cm
\hspace*{0in}\vspace*{.2in}
\epsffile{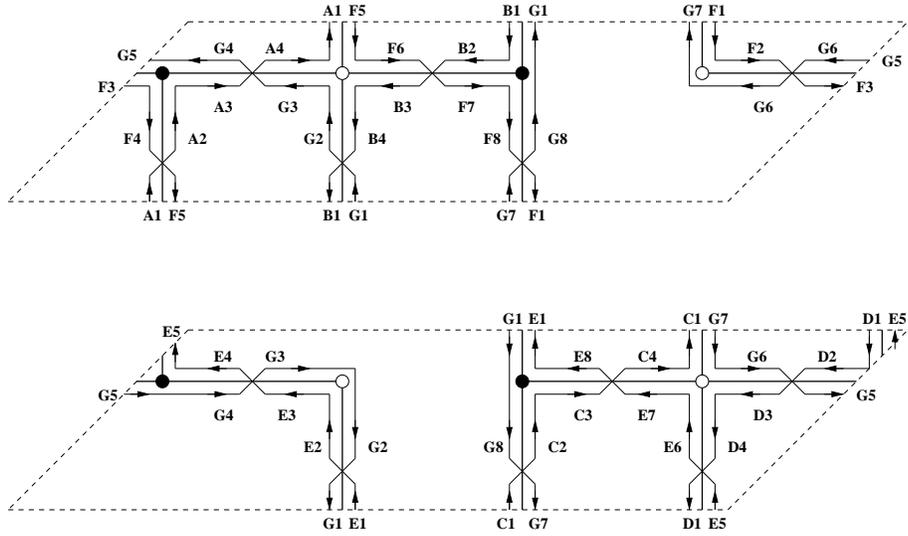}
\caption{\small Zig-zag paths corresponding to the two daughter theories, 
in the splitting of the double conifold singularity to two conifold 
singularities, with the corresponding dimers shown as thick lines.}
\label{dconidaug1}
\end{center}
\end{figure}

In Figure \ref{dconidaug1} we show the two sets of zig-zag paths. For 
convenience, the inherited paths are drawn in the locations corresponding 
to the original dimer. The information from the zig-zag paths allows to 
construct the dimer diagram corresponding to D3-brane at each of the 
left-over singularities after partial resolution. The dimer diagrams 
are also shown in the picture.

\begin{figure}
\begin{center}
\epsfxsize=10cm
\hspace*{0in}\vspace*{.2in}
\epsffile{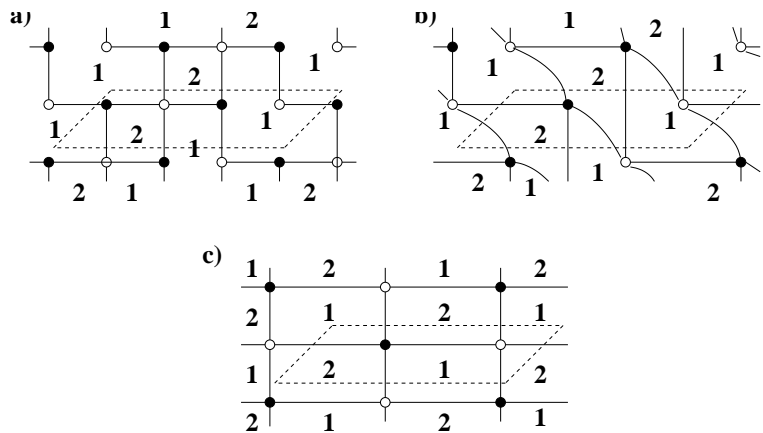}
\caption{\small (a) Dimer diagram corresponding to the first picture in 
Figure \ref{dconidaug1}. Figure (b) shows the dimer of the theory after 
integrating out massive modes. An equivalent diagram is shown in Figure 
(c), where one recognizes the dimer diagram of the conifold theory.}
\label{dconimass1}
\end{center}
\end{figure}

It is easy to convince oneself that the two theories are isomorphic (as 
expected from the symmetric factorization of the Riemann surface, or of 
the web diagram). Hence, it is enough to focus in one of them, say that 
shown in Figure \ref{dconimass1}a. Since this theory has a bi-valent 
node, 
one should integrate out the corresponding massive matter, with the result 
shown in Figure \ref{dconimass1}b. This can be redrawn as in Figure 
\ref{dconimass1}c, and one recognizes the dimer diagram for the conifold 
singularity, as expected. Hence the above technique of zig-zag paths 
provides a simple tool to determine the effect of a splitting by partial 
resolution on the dimer diagram of the D3-brane gauge theory, as a 
specific splitting of the initial dimer into two sub-dimers. Moreover, in 
section \ref{fieldtheory} we will show that the operation in the 
dimer diagram encodes in a 
simple manner the set of bi-fundamental vevs that corresponds in the gauge 
field theory to the partial resolution of the singularity.

The whole procedure can be subsumed in a simple operation in the dimer 
diagram, without the need to go through the Riemann surface.
In terms of the dimer diagram the previous discussion amounts to drawing
the old zig-zag paths that define the remaining singularity we are
interested in (C, D and E in the example above), and then
completing with new zig-zag paths (this will be G) until all edges
have two zig-zag paths going through them. The number of required new 
paths is given by the decrease in the number of holes in the factorization 
plus one. In the remaining examples we will obtain our results by using 
this simple shortcut. 

\subsection{Further examples and comments}
\label{moreexamples}

\subsubsection{Double conifold to two $\IC^2/\IZ_2$ singularities}

\begin{figure}
\begin{center}
\epsfxsize=10cm
\hspace*{0in}\vspace*{.2in}
\epsffile{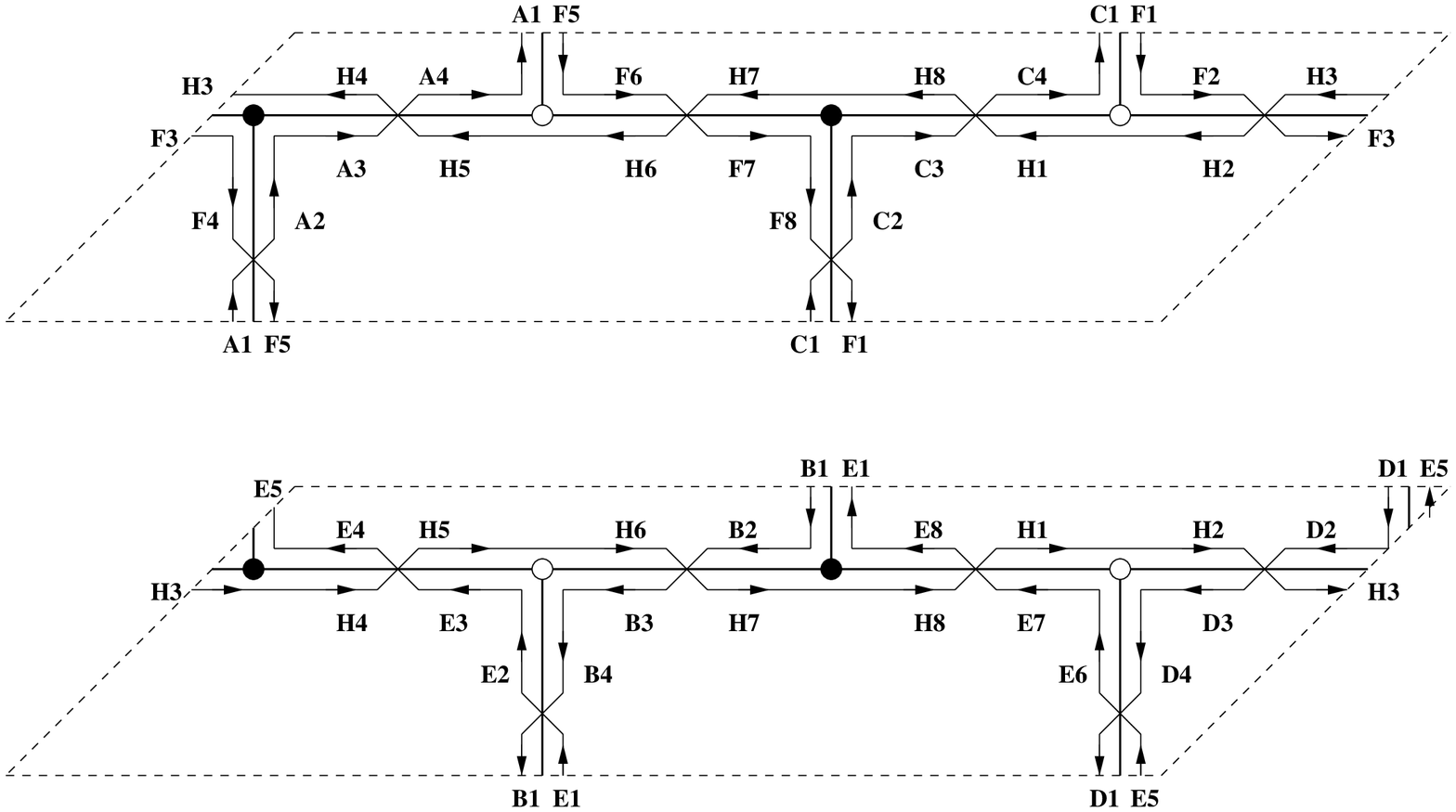}
\caption{\small Zig-zag paths corresponding to the two daughter theories, 
in the splitting of the double conifold singularity to two $\IC^2/\IZ_2$ 
singularities, with the corresponding dimers shown as thick lines.}
\label{dconidaug2}
\end{center}
\end{figure}

The technique we have described in the above example is fully general, and 
can be applied to any partial resolution. To provide an additional 
example, consider for instance the splitting of the above singularity 
into two $\IC^2/\IZ_2$ singularities, Figure \ref{dconisplit}b. 
Starting with the zig-zag paths in Figure \ref{dconipath}, the partial 
resolution corresponds to a factorization of the mirror Riemann surface 
splitting the set of paths into two subsets, namely A, C, F, and B, D, E.
Each set, along with a new path H from the new puncture in the daughter 
Riemann surface, allow to read off the dimer diagrams (and hence the 
quiver gauge theories) for D3-branes in the 
two left-over singularities. This is shown in Figure \ref{dconidaug2}, 
where one indeed recognizes the dimer diagrams of two $\IC^2/\IZ_2$ 
theories.

\subsubsection{From $dP_3$ to two SPP's}

\begin{figure}
\begin{center}
\epsfxsize=6cm
\hspace*{0in}\vspace*{.2in}
\epsffile{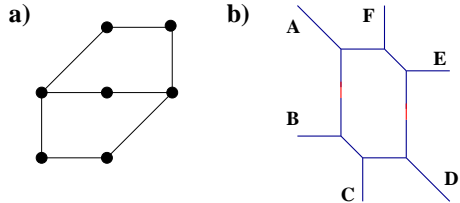}
\caption{\small The toric diagram and web diagram of the complex cone over 
$dP_3$, in a splitting to two SPP singularities. For clarity, the web 
diagrams of the left-over SPP singularities are shown for slightly 
resolved geometries. For future convenience, we have labeled external 
legs.}
\label{dp3toric}
\end{center}
\end{figure}

Before concluding this section, we present a further example, where the 
factorization lowers the genus of the mirror Riemann surfaces. Namely, the 
factorization implies elongating several segments in the web diagram.
Consider for instance the splitting of the complex cone over $dP_3$ to 
two suspended pinch point (SPP) singularities, shown in Figure \ref{dp3toric}.

\begin{figure}
\begin{center}
\epsfxsize=7cm
\hspace*{0in}\vspace*{.2in}
\epsffile{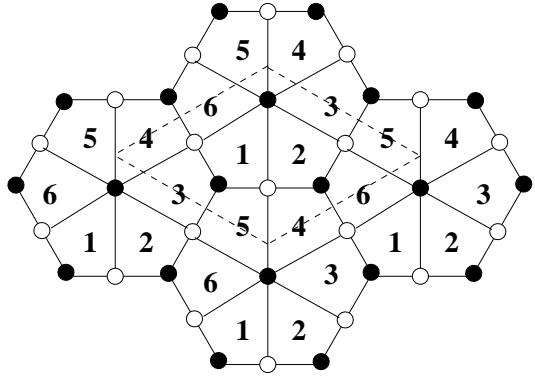}
\caption{\small 
The dimer diagram for the gauge theory of D3-branes 
on the complex cone over $dP_3$.}
\label{dp3dimer}
\end{center}
\end{figure}

\begin{figure}
\begin{center}
\epsfxsize=10cm
\hspace*{0in}\vspace*{.2in}
\epsffile{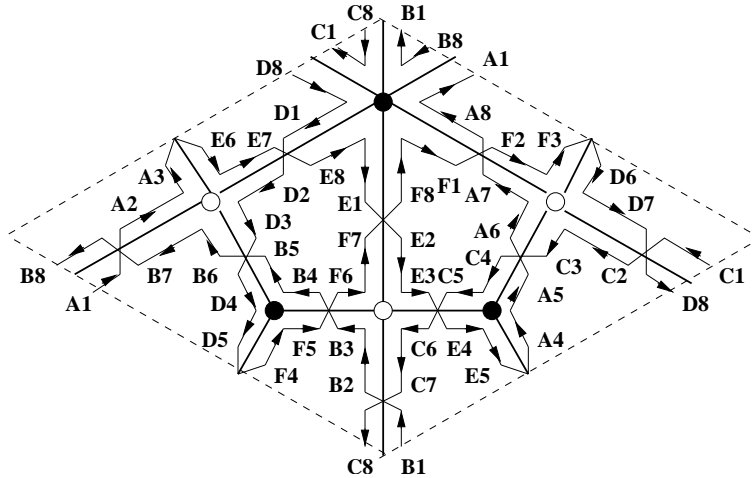}
\caption{\small The figure shows the unit cell of the dimer diagram for 
the gauge theory of D3-branes on the complex cone over $dP_3$, and 
the set of zig-zag paths.}
\label{dp3path}
\end{center}
\end{figure}

\begin{figure}
\begin{center}
\epsfxsize=16cm
\hspace*{0in}\vspace*{.2in}
\epsffile{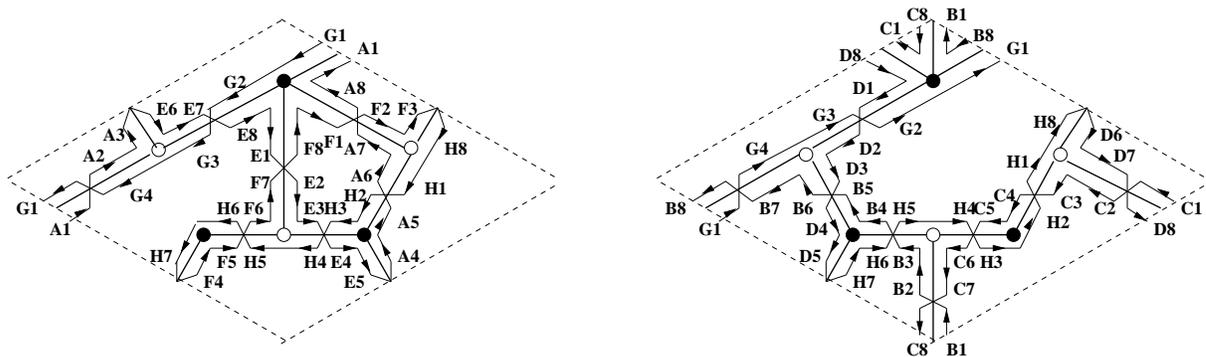}
\caption{\small The two dimers obtained upon the splitting by small 
resolution shown in Figure \ref{dp3toric}. They can be shown to be 
equivalent to two copies of the SPP dimer diagram.}
\label{dp3daug}
\end{center}
\end{figure}

The dimer diagram for (a toric phase of) the gauge theory on D3-branes at 
the cone over $dP_3$ is shown in Figure \ref{dp3dimer}.
The unit cell of the corresponding dimer diagram is shown in Figure 
\ref{dp3path}, where we also show the zig-zag paths. The 
partial resolution in Figure \ref{dp3toric} has the effect of splitting 
this dimer diagram into the two dimer diagrams in Figure \ref{dp3daug}. 
After integrating out massive fields, they can be shown to correspond to 
the gauge theories of D3-branes at SPP singularities, in agreement with 
the underlying geometric picture. Although this example follows from 
exactly the same rules as previous ones, we encounter the new feature that 
the splitting of the dimer involves two new zig-zag paths (denoted G and 
H) rather than one. This simply reflects the fact that the factorization 
of the Riemann surface involves two elongated tubes, hence two new 
punctures for each daughter Riemann surface.

\subsubsection{Minimal partial resolution}
\label{minimalresolution}

To conclude this section, we would like to mention that this technique can 
be applied to asymmetric splittings, where the two daughter geometries are 
not the same. One particular extremal case is a minimal partial resolution 
(removing only one triangle from the toric diagram). Hence, only one 
singularity is left over after the partial resolution (namely the second 
singularity turns out to be a smooth patch). Let us describe this more 
explicitly.

In terms of the web diagram, this  simply corresponds to elongating a tube 
that separates two external legs from the rest of the web. Using the 
zig-zag paths, it is easy to show that the left-over singularity 
corresponds to a dimer diagram obtained from the initial one by the 
removal of some edges. These edges are precisely those over which the two 
zig-zag paths associated to the removed legs overlap \footnote{This 
description explains as in \cite{Hanany:2005ss} the possibility of the 
appearance of inconsistent 
dimer diagrams by arbitrary addition/removal of edges. For instance, 
consider a minimal partial resolution involving two zig-zag paths 
overlapping over more than one edge. The removal on only one of these 
edges does not correspond to a consistent separation of zig-zag paths and 
leads to an inconsistent diagram.}

To provide one particular example, we describe the partial resolution of 
the double conifold to an SPP singularity via the removal of one triangle 
in the corresponding toric diagram. Concretely consider separating the 
legs A and F from the rest of the web diagram, by stretching the 
intermediate segment. Since the corresponding zig-zag paths overlap over 
the lower left edge in the dimer diagram, this is the edge to be removed.
In field theoretic terms this means that the corresponding
bifundamental gets a vev, and the two faces (gauge groups) sharing the
edge join (gauge factors break to the diagonal combination). The resulting 
dimer diagram is that of the SPP theory, as can be checked by computing 
the gauge theory data.

We hope these examples suffice to illustrate the general validity of the 
above prescription.

\subsection{Field theory interpretation}
\label{fieldtheory}

As discussed in \cite{Morrison:1998cs}, partial resolutions of 
singularities correspond to turning on Fayet-Iliopoulos terms on the gauge 
theory on D3-branes sitting at them. These FI terms force some of the 
bi-fundamental scalars to acquire vevs, preserving supersymmetry but 
partially breaking gauge symmetry, in precise agreement with the quiver
gauge theory on D3-branes at the final left-over singularity.

In this section we show that the operation of splitting a dimer, as 
described in previous section, encodes in a very precise fashion 
the field theory data corresponding to the Higgs mechanism and gauge 
symmetry breaking. Moreover we show that dimer techniques can be 
efficiently used to show the F- and D-flatness of such vevs.

For simplicity, we center on a gauge theory with all gauge factors having 
equal rank $N$. Discussion of other situations (fractional branes) is 
postponed until section \ref{fractional}. We also consider that after the 
splitting, $N_1$ D3-branes remain at the first singularity and $N_2$ 
remain at the second.

In order to describe the bi-fundamental vevs in the field theory, we 
notice that in the dimer splitting there are three 
different kinds of bi-fundamental fields, according to the behaviour of 
the corresponding edge: a) those appearing in the two 
daughter dimers; b) those not appearing in the first sub-dimer, but 
present in the second; c) those not appearing in the second, but 
present in the first. This suggests the following ansatz for their vevs, 
which we denote $V_0$, $V_1$, $V_2$, respectively:
\beqa
V_0 \, = \, \pmatrix{0 & 0 \cr 0 & 0} \quad ; \quad
V_1 \, =\, \pmatrix{v\, \id_{N_1} & 0 \cr 0 & 0} \quad ; \quad
V_2 \, =\, \pmatrix{0 & 0 \cr 0 & v\, \id_{N_2} } 
\label{vevansatz}
\eeqa
where bi-fundamental fields are regarded as $N\times N$ matrices, and the 
entries are blocks of dimension appropriate to the partition $N=N_1+N_2$.
Here we take $v$ to be adimensional, and we consider that a dimensionful 
constant enters into the vev of each bi-fundamental, exponentiated to the 
appropriate power to match its conformal dimension. This factor does 
not change the discussion of flatness, hence we ignore it in what follows.

The interpretation of this ansatz is very clear. The $N_1$, $N_2$ entries 
in the diagonal determine the pattern of gauge symmetry breaking triggered 
by that bi-fundamental for the set of the $N_1$, $N_2$ D3-branes in the 
first, resp. second dimer. An edge absent in a sub-dimer implies
a local recombination of the corresponding set of D3-branes across the 
associated bi-fundamental. Namely, there is a non-vanishing vev in the 
corresponding set of entries. Similarly, for edges present in a sub-dimer 
there is no vev in the corresponding entries of the associated 
bi-fundamentals.

The proof that the above assignment of vevs satisfies the flatness 
conditions in the field theory is provided in appendix \ref{proof}. 
However it is useful to work out an explicit example, so consider for 
instance the splitting of the 
double conifold to two conifold singularities. Using the information in 
Figure \ref{dconidimer} for the initial dimer, and Figure 
\ref{dconidaug1} for the sub-dimers, we obtain the following set of vevs
\beqa
\Phi_{12}\, =\, V_2 \quad , \quad \Phi_{23}\, =\, V_2 \quad , \quad
\Phi_{34}\, =\, V_0 \quad , \quad \Phi_{41}\, =\, V_0 \nonumber\\
\Phi_{21}\, =\, V_0 \quad , \quad \Phi_{32}\, =\, V_0 \quad , \quad
\Phi_{43}\, =\, V_1 \quad , \quad \Phi_{14}\, =\, V_1 
\label{thevevs}
\eeqa 
where we have introduced the notation $\Phi_{ij}$ for a bi-fundamental 
$(\fund_i,\antifund_j)$, and take $\Phi_{12}$ to correspond to the 
vertical edge in the left part of the depicted unit cell.

It is now straightforward to analyze the flatness conditions on the set 
of vevs for this example.
Concerning the F-term conditions, all nodes are 4-valent, hence the 
superpotential is a sum of quartic terms. Moreover, any such term
contains at least two fields without vev. Hence, the F-terms conditions 
are automatically satisfied. Concerning the non-abelian D-term conditions, 
we write the generators of $SU(N)$ as
\beqa
T=\pmatrix{T_{11} & T_{12} \cr T_{21} & T_{22}}
\eeqa
and obtain that the D-term contributions for the $SU(N)_i$ factors are
\beqa
& SU(N)_1 \quad \quad \tr(\Phi_{12}^\dagger T \Phi_{12}) + 
\tr(\Phi_{14}^\dagger T \Phi_{14})  \, =\, |v|^2\, ( \tr T_{11} + \tr T_{22})\, 
=\, 0 \nonumber \\
& SU(N)_2 \quad \quad -\tr(\Phi_{12}^\dagger T \Phi_{12}) + 
\tr(\Phi_{23}^\dagger T \Phi_{23})  \, =\, |v|^2\, ( \tr T_{22}- \tr T_{22})\, 
=\, 0 \nonumber \\
& SU(N)_3 \quad \quad -\tr(\Phi_{23}^\dagger T \Phi_{23}) - 
\tr(\Phi_{43}^\dagger T \Phi_{43})  \, =\, -|v|^2\, ( \tr T_{11} + \tr T_{22})\, 
=\, 0 \nonumber \\
& SU(N)_4 \quad \quad \tr(\Phi_{43}^\dagger T \Phi_{43}) -
\tr(\Phi_{14}^\dagger T \Phi_{14})  \, =\, |v|^2\, ( \tr T_{11}- \tr T_{11})\, 
=\, 0
\label{dflat}
\eeqa
where we have used tracelessness of $SU(N)$ generators. Finally, 
concerning the abelian D-term conditions, the above vevs lead to non-zero 
contributions which are suitably canceled by the non-zero FI terms. This 
effective absence of $U(1)$ D-term constraints can be equivalently 
regarded as the statement that there are $B\wedge F$ couplings (related to 
the FI terms by supersymmetry) which render the $U(1)$'s massive, so that 
they are not present at low energies and hence no D-term constraints have 
to be imposed, see footnote \ref{uunos}.

Notice that the description in this section generalizes in a 
straightforward fashion to the splitting of a dimer into more than two 
sub-dimers.

\subsection{Effect on perfect matchings}
\label{matchings}

\begin{figure}
\begin{center}
\epsfxsize=12cm
\hspace*{0in}\vspace*{.2in}
\epsffile{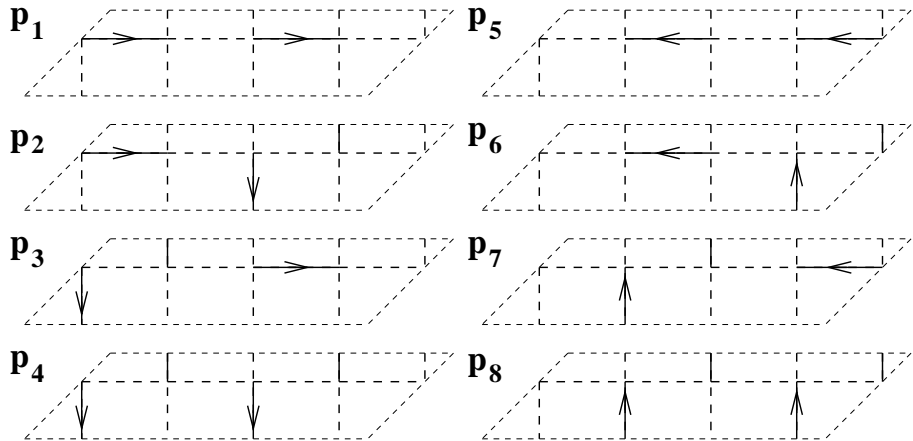}
\caption{\small The eight perfect matchings for the dimer diagram of the 
double conifold.}
\label{dconipmatch}
\end{center}
\end{figure}

\begin{figure}
\begin{center}
\epsfxsize=10cm
\hspace*{0in}\vspace*{.2in}
\epsffile{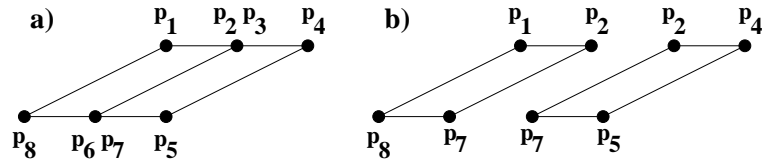}
\caption{\small In a partial resolution, the original perfect matchings 
descending to perfect matchings of one or the other subdimer end up 
located at one or the other toric sub-diagram, as shown here for the 
resolution of the double conifold to two conifolds.}
\label{dconidescend}
\end{center}
\end{figure}

It is interesting to consider the effect of partial resolution on perfect 
matchings. This can be easily analyzed at the level of the dimer diagrams, 
as we do in what follows in a particular example. Consider the double 
conifold, whose dimer diagram is shown in Figure \ref{dconidimer}. 
The eight perfect matchings for this diagram are shown in Figure 
\ref{dconipmatch}. The location of these matchings in the toric diagram, 
obtained as described in Section \ref{pmatchings}, using $p_1$ as 
reference matching, are shown in Figure \ref{dconidescend}a.

Consider the partial resolution of the double conifold to two conifolds, 
studied in Section \ref{dconi}, whose two resulting sub-dimers are 
shown in Figure \ref{dconidaug1}. The splitting of the dimer into 
sub-dimers implies that the perfect matchings of the original dimer fall 
into different classes: 
\begin{itemize}

\item The perfect matchings $p_4,p_5$ descend to perfect matchings of the 
first sub-dimer.

\item The perfect matchings $p_1,p_8$ descend to perfect matchings of the 
second sub-dimer.

\item The perfect matchings $p_2,p_7$ correspond to perfect matchings of 
both the first and second sub-dimer.

\item The perfect matchings $p_3,p_6$ do not correspond to perfect 
matchings of either sub-dimer.

\end{itemize}

This correspondence becomes nicely meaningful when one considers the 
location of the different perfect matchings in the toric diagram. The 
partial resolution splits the toric diagram in two pieces, separated by a 
common internal segment. Perfect matchings of the original dimer which 
descend to perfect matchings of a given sub-dimer are located at
points on the piece of the toric diagram describing the corresponding
daughter singularity.
Perfect matchings descending to matchings of both singularities are 
located along the common segment in the toric diagram. This is described 
for the double conifold in Figure \ref{dconidescend}.

It is possible to show that this pattern is completely general, and that 
for a general partial resolution perfect matchings fall into one of these 
four classes. Namely, we label the edges of the dimer diagram with 
labels 1, 2 and 3, according to whether it is present in sub-dimer one, or 
in sub-dimer 2, or in both. Perfect matchings involving edges of type 1 
and 3 end up in the interior of the toric sub-diagram 1; perfect matchings 
involving edges of type 2 and 3 end up in the interior of the toric 
sub-diagram 2; perfect matchings only involving edges of type 3 appear on 
both toric sub-diagrams, along their common boundary; perfect matchings 
with edges of type 1 and 2 (and possibly 3) disappear. 

\medskip

One can also obtain the effect of the partial resolution on the perfect 
matchings from the viewpoint of the Riemann surface. For that, one can use 
the relation described in Section \ref{pmatchings} between pairs of 
perfect matchings and 1-cycles on the mirror Riemann surface. The first 
observation is that a partial resolution corresponds to the introduction 
of a segment joining two external non-adjacent perfect matchings 
$p$, $p'$ in the toric diagram. This is just dual to separating the web 
diagram by elongating the leg dual to that segment. Notice that cases 
where there are multiple matchings at the 
corresponding points in the toric diagram simply correspond to cases where 
there are several parallel legs in the web diagram, and correspondingly 
several possibilities to perform the partial resolution. For instance, in 
our above example, the partial resolution corresponds to choosing the
perfect matchings $p_2$ and $p_7$.

To such a pair of perfect matchings one can associate a path 
$p'-p$ in the 
dimer diagram and a 1-cycle in the mirror Riemann surface. In fact, this 
1-cycle wraps around the tube which becomes infinitely elongated in the 
partial resolution process. In terms of the dimer diagram, it means that 
the path in the dimer diagram becomes the new zig-zag path 
(denoted G in our example in Section \ref{dconi}) introduced to 
construct the new sub-dimers.

Given that this 1-cycle separates the Riemann surfaces in two pieces, 
which are naturally associated to the two daughter singularities, it is 
possible to interpret the four classes of perfect matchings 
in terms of their behaviour on the Riemann surface $\Sigma$. Consider one 
of the external perfect matchings e.g. $p$. For any other matching $p_i$ 
one can consider the 1-cycles associated to $p_{i}-p$ obtained using the 
tiling of $\Sigma$. If the whole of such 1-cycle lies on one component of 
$\Sigma$, the matching $p_i$ will correspond to a perfect matching of the 
corresponding sub-dimer, and to a point in the corresponding toric  
sub-diagram. If all 
pieces of the 1-cycle are included in the 1-cycle $p'-p$, then $p_i$ will 
correspond to a perfect matching of both sub-dimers, and will appear in 
both toric diagrams (concretely, along the common boundary). Finally if 
the 1-cycle contains pieces lying in both components of $\Sigma$, the 
corresponding perfect matching disappears in the process of partial 
resolution.

These properties are easily explicitly checked in our above example, and 
can be generalized to any partial resolution. We leave the discussion as 
an exercise for the interested reader.

\subsection{Partial resolutions with fractional branes}
\label{fractional}

In this section we would like to study partial resolutions for 
singularities in the presence of fractional branes, and their description 
using dimers. For concreteness we center on a particular example, although 
our conclusions are of general validity.

\begin{figure}
\begin{center}
\psfrag{N}{$N$}
\psfrag{NplusM}{$N+M$}
\psfrag{NplusQ}{$N+Q$}
\psfrag{NplusP}{$N+P$}
\psfrag{N1}{$N_1$}
\psfrag{N1plusM1}{$N_1+M_1$}
\psfrag{N2}{$N_2$}
\psfrag{N2plusM2}{$N_2+M_2$}
\includegraphics[scale=.6]{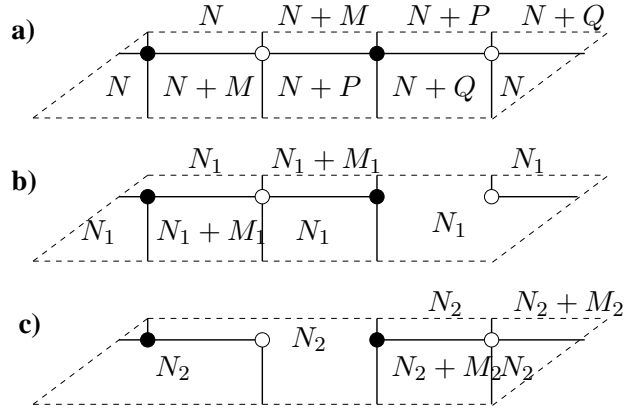}
\caption{\small a) The dimer for the double conifold with the most general 
set of fractional branes. b,c) The sub-dimers for the daughter conifold 
singularities, with their fractional branes.
}
\label{dconifract}
\end{center}
\end{figure}

Let us consider the splitting of the double conifold to two conifold 
singularities. The dimer diagram for the double conifold, with the most 
general set of fractional branes, is shown in Figure \ref{dconifract}a. 
Since the field theory is non-chiral, there are no restrictions on the 
gauge factor ranks, and hence there are three kinds of fractional branes.

When the singularity is split into two conifolds, the latter may contain 
fractional branes as well. The most general possibility is shown in Figure 
\ref{dconifract}b, c. Since each conifold allows for one kind of 
fractional brane, there are two possible fractional branes in the final 
system.

It is thus a natural question to ask what happens with the third kind of 
fractional brane. The answer, that we can recover from different 
viewpoints, is that it obstructs the partial resolution. A 
pictorial way to derive this result is to compare the original dimer and 
the daughter sub-dimers in Figures \ref{dconifract}a and b,c, 
respectively. In order to have a proper splitting, the number of branes in 
a given face of the original dimer must agree with the sum of the numbers 
of branes in the corresponding location in the sub-dimers. In our 
particular case, this implies
\beqa
N=N_1+N_2 \quad , \quad M=M_1 \quad , \quad P=0 \quad , \quad Q=M_2
\eeqa
Hence we see that the splitting necessarily forces the fractional brane 
changing the rank of the gauge group $3$ to be absent, in the sense that 
only in the absence of such brane the splitting is possible. More 
precisely, what obstructs the splitting is the fractional brane which 
controls the difference between the ranks of the gauge factors 1 and 3.

In what follows we present several interpretations for this fact. From the 
viewpoint of the field theory of the initial singularity, it means that 
the theory with different ranks for the factors 1 and 3 does not have the 
corresponding flat direction. This can be argued in general, but is 
suffices to discuss one particular example, for instance $M=Q=0$, 
$P\neq 0$. It is simple to show that the D-term conditions for gauge 
factor 3 cannot be satisfied. Indeed, the natural ansatz is similar to 
(\ref{thevevs}), with the only difference that for non-square matrices, 
the entries in the $M\times P$ additional submatrix are taken to be zero.
In computing the D-term, as in (\ref{dflat}), for the gauge factor 3, one 
notices that the non-zero vevs do not suffice to complete the full 
$SU(M+P)$ trace, and hence the D-term does not vanish.

One can regard the dimer as realizing a physical construction of the gauge 
theory in terms of NS-branes and D5-branes \cite{Franco:2005rj}, similar 
to brane box \cite{Hanany:1997tb,Hanany:1998ru,Hanany:1998it}
or brane diamond models \cite{Aganagic:1999fe}. In that context, the relation 
between obstructions to splitting the brane configuration and D-terms in 
the gauge theory is familiar and well-known. Intuitively, the motion of 
the NS-branes drags a subset of the D5-branes, increasing their tension 
and breaking supersymmetry (equivalently, misaligning the phase of their 
BPS charge with respect to the others). In the absence of fractional 
branes, D5-branes on opposite sides of the NS-brane can recombine and 
snap back, restoring supersymmetry at the price of breaking gauge 
symmetry (equivalently, forming a bound state with appropriately aligned 
phase). For certain fractional branes the possibility of recombination is
not possible, leading to an obstruction to the brane motion.

\begin{figure}
\begin{center}
\epsfxsize=5cm
\hspace*{0in}\vspace*{.2in}
\epsffile{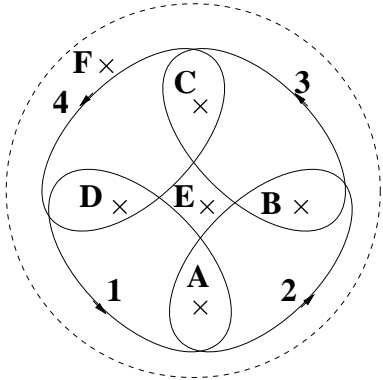}
\caption{\small The 1-cycles in $\Sigma$ corresponding to the D-branes 
controlling the rank of the different gauge factors in the double conifold 
gauge theory.}
\label{dconi3cycles}
\end{center}
\end{figure}

To conclude this section, we provide an interpretation of the obstruction 
in terms of the mirror geometry, where a very explicit version of the 
above picture can be derived. In order to do that, consider the 1-cycles 
on the mirror Riemann surface $\Sigma$ which correspond to the different 
fractional branes, in our case, to the different faces in the dimer. 
These are sketched in Figure \ref{dconi3cycles}. 

The structure of these 1-cycles, and in 
particular their winding around the punctures of $\Sigma$, leads to a 
natural explanation of the obstruction. Consider introducing only 
fractional branes changing the rank of the gauge factor 3. In the 
mirror this corresponds to introducing D-branes along the cycle that 
surrounds the punctures B, C. These punctures end up in different 
daughter Riemann surfaces in the splitting, see figure \ref{dconifactor}, 
hence in trying to perform 
the partial resolution, the D-brane stretches along the elongated 
tube, hence increases its tension and breaks supersymmetry. Moreover, it 
is not possible to express this 1-cycle in terms of a combination of 
brane cycles not stretching along the tube, hence no process 
restoring supersymmetry can take place. The same argument goes through if 
one considers only fractional branes of type 1, since they surround the 
punctures A, D. Notice that there is no problem if one considers instead
fractional branes of type 2 or of type 4, since they do not correspond to 
cycles stretching along the tube.

Finally, consider introducing the same number of fractional branes of type 
1 and 3. This case leads to equal rank for gauge factors 1 and 3, and 
hence we expect no obstruction. Indeed, although the brane correspond to 
cycles stretching along the tube, it is possible to deform them 
topologically to a sum of cycles of type 2 and 4, which do not stretch. 

As mentioned above, this picture generalizes to more involved situations. 
The general lesson is that sets of fractional branes associated 
to cycles stretching along the tubes which elongate in the factorization 
of the Riemann surface lead to obstructions to the partial resolution.

\section{Complex deformations}
\label{complexdeformations}

In this section we discuss the smoothing of singular geometries via a 
different process, namely complex deformations. Again, our analysis is 
general and valid for complex deformations which partially smooth out a 
singularity, or which split it into two daughter singularities.

Complex deformations of toric singularities have been discussed in diverse 
contexts. They are easily characterized in terms of the web diagram, as a 
splitting of the web into two sub-webs in equilibrium. Pictorially, the 
segment suspended between the two sub-webs after splitting represents the 
3-cycle in the deformed geometry. This description, phrased in terms of a 
physical realization as a fivebrane web in 
\cite{Aharony:1997ju,Aharony:1997bh} and in toric language in 
\cite{Leung:1997tw,Aganagic:2001ug}, is actually 
based on the mathematical theory of complex deformations 
\cite{altmann1,altmann2,altmann3}.

In what concerns the relation between complex deformations and the gauge 
theory on D3-branes, the situation is very different from partial 
resolutions, and has been studied in \cite{Franco:2005fd} (generalizing 
the conifold case in \cite{Klebanov:2000hb}). The D-branes 
always live in the resolution phase of the singular geometry. Complex 
deformation phases are realized after geometric transitions triggered by 
the back-reaction of a large number of fractional branes \footnote{Here 
by fractional brane we mean any anomaly-free assignment of ranks to the 
quiver gauge theory, which does not correspond to all factors having 
equal rank. Also it is important to point out that we center on the case 
of `deformation branes' in the sense of \cite{Franco:2005zu}, since there 
exist other kinds of fractional 
branes with different dynamical behaviour, like the removal of 
supersymmetric vacuum by strong infrared effects 
\cite{Berenstein:2005xa,Franco:2005zu,Bertolini:2005di}, see also
\cite{Intriligator:2005aw}.}. Namely, the homology class wrapped by the 
fractional brane disappears, and it is replaced by a 3-cycle in a complex 
deformation phase, which supports 3-form fluxes. This is similar to the 
Klebanov-Strassler proposal for the conifold with fractional branes 
\cite{Klebanov:2000hb}.

The description of `deformation' fractional branes in terms of the dimer 
diagrams was provided in \cite{Franco:2005zu}. Moreover a simple 
procedure was suggested to transform the original dimer diagram into 
the dimer diagram of the singular geometry after a complex deformation. 
Deformation branes correspond to a clusters of faces, touching at their 
corners, and complex deformation has the effect of smoothing certain 
touching edges and collapsing the painted regions to zero size.
Although efficient, this procedure was rather {\em{ad hoc}}, with no 
derivation from first principles, and no clear rules on which 
intersections to smooth out, etc. In this section we fill this gap, and 
find clear rules based on physical principles of geometric transitions.

\subsection{An example in the double conifold}
\label{dconicomp}

Let us start by considering a simple example, namely the complex 
deformation of the double conifold singularity into a conifold. In order 
to emphasize the physical ideas, we start with the gauge theory 
description and derive the effect on the dimer and the description in the 
web diagram.

\begin{figure}
\begin{center}
\epsfxsize=14cm
\hspace*{0in}\vspace*{.2in}
\epsffile{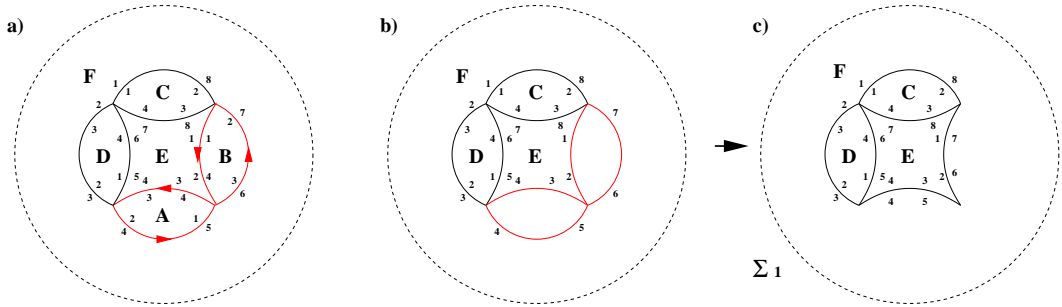}
\caption{\small The geometric transition in the mirror picture. (a) the 
fractional brane defines a 1-cycle in the Riemann surface, along which we 
cut it (b). Gluing the boundaries of the later gives a daughter Riemann 
surfaces encoding the mirror geometry of the left over geometry (c).}
\label{dconibreak2}
\end{center}
\end{figure}

Consider the gauge theory for D3-branes at the double conifold singularity 
(whose dimer diagram is in Figure \ref{dconidimer}) in the presence of 
additional fractional branes, of the kind that increase the rank of the 
gauge factor 2. The geometric transition triggered by fractional 
branes is most easily seen in the mirror picture, in terms of an operation 
on the mirror Riemann surface. In figure \ref{dconibreak2}a we show the 
mirror Riemann surface, along with the 1-cycle on which the corresponding 
fractional brane wraps. It corresponds to a 1-cycle winding around the 
punctures labeled A and B. The geometric transition corresponds to cutting 
the Riemann surface along this 1-cycle, as shown in Figure 
\ref{dconibreak2}b, and gluing the boundary to get a new Riemann 
surface $\Sigma_1$, as shown in Figure \ref{dconibreak2}c. 

Notice that in fact this process leads to to two disjoint Riemann 
surfaces. The second (denoted $\Sigma_2$) corresponds to the removed 
pieces A, B, with a suitable closing of its boundary. In cases where the 
1-cycle splits off two punctures, like in our example, this second 
Riemann surface is somewhat degenerate and we do not shown it in the 
pictures.

\begin{figure}
\begin{center}
\epsfxsize=8cm
\hspace*{0in}\vspace*{.2in}
\epsffile{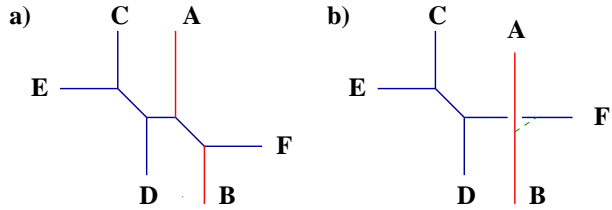}
\caption{\small Complex deformation of the double conifold to a conifold 
singularity.}
\label{dconicomplex}
\end{center}
\end{figure}

The distance between the two Riemann surfaces in the ambient space 
corresponds to the size of the new cycle, which is a 3-cycle in the 
original D3-brane configuration. The 1-cycle on which the fractional 
D-branes were wrapping has disappeared as required.
The process can be regarded as a splitting of the original web diagram 
into two sub-webs in equilibrium, given by the sets C, D, E, F and A, B 
respectively. Hence, from the physics of geometric 
transitions we recover the usual description in terms of the web diagram 
as shown in Figure \ref{dconicomplex}. A bonus of our argument is that it 
allows a direct identification of which web splitting corresponds to the 
geometric transition triggered by a given fractional brane. This piece of 
the dictionary was missing from previous analysis \cite{Franco:2005fd}, 
which supplemented it with suitable guesswork.

\begin{figure}
\begin{center}
\epsfxsize=10cm
\hspace*{0in}\vspace*{.2in}
\epsffile{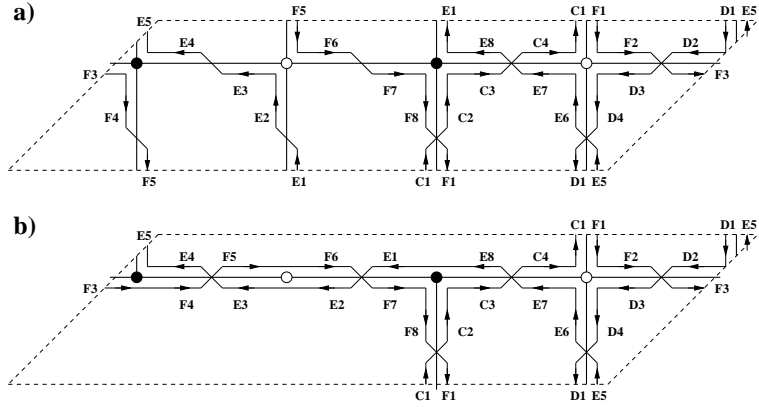}
\caption{\small 
a) Zig-zag paths of the original dimer corresponding to the punctures in 
$\Sigma_1$. b) The closing of the open boundary to get $\Sigma_1$ implies 
adjacency relations that allow to reconstruct the edges of the dimer 
diagram associated to the D3-branes in the complex deformed geometry. In 
this case we recover the dimer diagram of a conifold, as expected from a 
complex deformation of the double conifold.}
\label{dconidimdef}
\end{center}
\end{figure}

The new Riemann surfaces moreover allow us to construct the dimer theories 
of the theories in the left over geometry after complex deformation. 
Namely, for each of the Riemann surfaces, the left 
over zig-zag paths, along with the adjacency relations (including those 
required by the gluing of boundaries in figure \ref{dconibreak2}), can be 
used to define the left-over dimer diagram \footnote{For Riemann surfaces 
with two punctures the dimer diagram is somewhat peculiar, with one face, 
one edge and no nodes. The corresponding gauge theory can however be 
properly obtained by recalling the physical realization of the dimer in 
terms of NS- and D5-branes. The special rules for this diagram are due to 
the extended supersymmetry of the configuration.}. This is shown in 
Figure \ref{dconidimdef} for $\Sigma_1$. In figure \ref{dconidimdef}a we 
depict the zig-zag paths of the original dimer diagram which are 
associated to punctures in the daughter Riemann surface $\Sigma_1$. In 
Figure \ref{dconidimdef}b we implement the new adjacency relations 
implied by the gluing of the boundary in Figure \ref{dconibreak2}c. We 
also show the dimer diagram that corresponds to this final configuration. 
After integrating out the matter massive due to the bi-valent node, the 
diagram is easily shown to correspond to the conifold theory, hence 
describes the gauge theory on D3-branes sitting at the left 
over conifold singularity after complex deformation.

\begin{figure}
\begin{center}
\epsfxsize=14cm
\hspace*{0in}\vspace*{.2in}
\epsffile{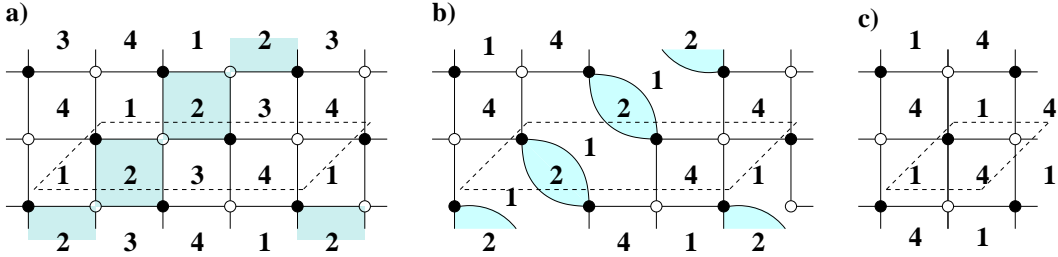}
\caption{\small Effect of the complex deformation on the dimer diagram, 
according to the rules in \cite{Franco:2005fd}. (a) shows the fractional 
branes upon consideration. In (b) the painted areas are contracted, and 
the nodes where they touched are split open. In (c) we show the 
contraction of painted regions is completed, leaving a final dimer diagram 
corresponding to the conifold theory.}
\label{dconiadhoc}
\end{center}
\end{figure}

Notice that this operation on the dimer is equivalent to the proposal in 
\cite{Franco:2005fd} of collapsing some faces to zero, while splitting 
some nodes open, shown in Figure \ref{dconiadhoc}. The key difference is 
that in our analysis the rules are derived from first principles
thanks to a proper understanding of the geometric transition and its
implications on the dimer.

\subsection{An example in $dP_3$}
\label{moreexamplescomplex}

Let us consider another example, in fact one of the complex deformations 
discussed in \cite{Franco:2005fd}. Consider the gauge theory on 
D3-branes at the complex cone over $dP_3$, whose dimer diagram is shown 
in Figure \ref{dp3dimer}, with a number of fractional branes increasing 
the rank of the gauge factors 1, 3 and 5 by the same amount. The gauge 
theory analysis in \cite{Franco:2005fd} suggests that these fractional
branes trigger a complex deformation smoothing the singularity completely. 

\begin{figure}
\begin{center}
\epsfxsize=8cm
\hspace*{0in}\vspace*{.2in}
\epsffile{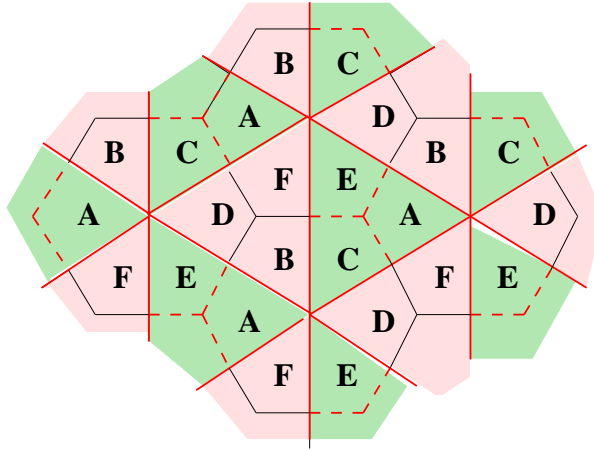}
\caption{\small The mirror Riemann surface for $dP_3$ is a torus with six 
punctures, which is depicted as a periodic array (the fact that it 
formally looks like the original dimer diagram is accidental, and not 
true for a general singularity). The total 1-cycle that 
corresponds to the fractional brane in the discussion corresponds to a 
triangular path enclosing the punctures A, 
C, E, and is shown in red. It separates the Riemann surface in two 
pieces, shown in different color, each with the topology of a disk.} 
\label{dp3break}
\end{center}
\end{figure}

\begin{figure}
\begin{center}
\epsfxsize=8cm
\hspace*{0in}\vspace*{.2in}
\epsffile{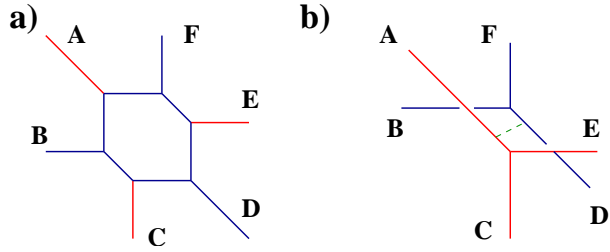}
\caption{\small Complex deformation of the cone over $dP_3$ to a 
smooth space. The sub-webs describing the left over singularities 
actually describe two copies of (locally) flat space.}
\label{dp3complex}
\end{center}
\end{figure}

We can easily recover this result following the procedure described above 
in the dimer diagrams. Thus one constructs the 1-cycle in the mirror 
Riemann 
surface that is the homological sum of the 1-cycles corresponding to the 
faces 1, 3 and 5. Working this out as in the above example, this total 
1-cycle winds around the punctures A, C and E, namely separates the 
Riemann surface in two regions, which include the punctures A, C, E and B, 
D, F, respectively. This is shown in figure \ref{dp3break}. The geometric 
transition triggered by the fractional 
branes should make this cycle disappear, hence we cut the Riemann surface 
along this cycle, and glue the boundary of each to yield two daughter 
Riemann surfaces. Each has the topology of a 2-sphere with three 
punctures. Clearly this process corresponds to a splitting of the web 
diagram in two sub-webs, in particular to the splitting shown in Figure 
\ref{dp3complex}. This had already been obtained in \cite{Franco:2005fd} 
by guesswork.

\begin{figure}
\begin{center}
\epsfxsize=14cm
\hspace*{0in}\vspace*{.2in}
\epsffile{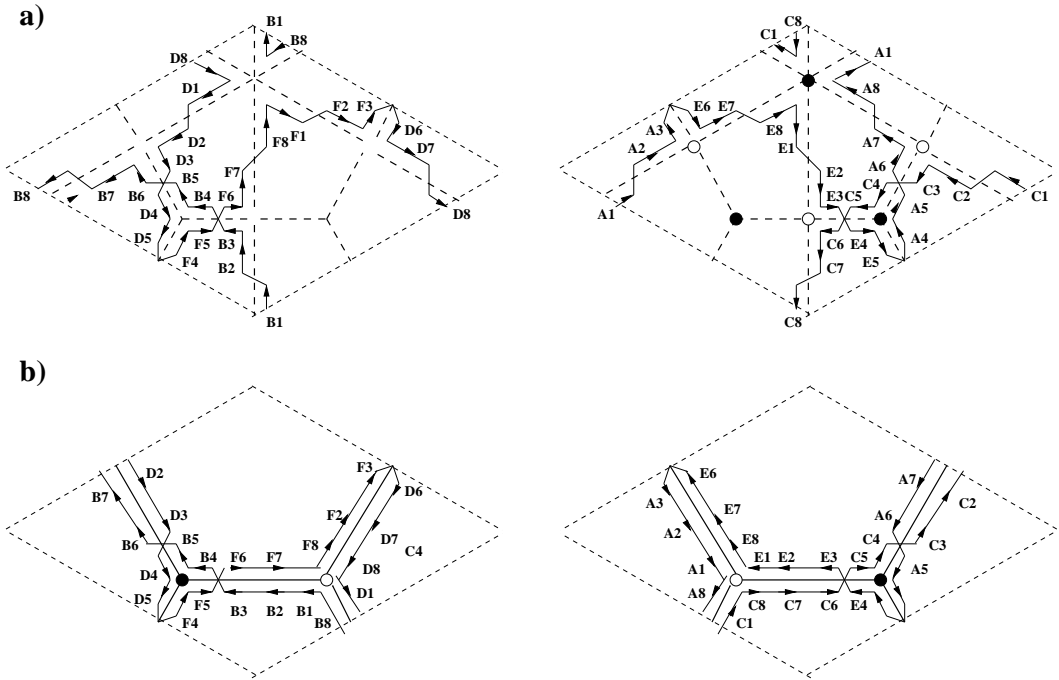}
\caption{\small The effect of the complex deformation on the dimer diagram 
of $dP_3$. Figure a) shows the two sets of zig-zag paths corresponding to 
the two pieces of Riemann surface obtained after cutting along the 1-cycle 
defined by the fractional brane discussed in the main text. Figure b) 
shows the new zig-zag paths obtained upon imposing the adjacency relations 
implied by the gluing of boundaries of the Riemann surfaces. Notice these 
adjacency relations are completely natural from the viewpoint of zig-zag 
paths; they simply amount to pairing unpaired pieces in a consistent way.} 
\label{dp3dimdef} 
\end{center} 
\end{figure}

The above operations in the 1-cycles in the mirror Riemann surface, have a 
direct effect on the set of zig-zag paths of the original theory. In 
fact, the resulting dimers after the complex deformation can be recovered by 
directly operating on these. Consider the zig-zag paths of the original 
$dP_3$ theory, shown in Figure \ref{dp3path}. The cutting of the Riemann 
surface separates them into two daughter sets, that will correspond to the 
zig-zag paths of the daughter theories. They are shown in the 
Figure \ref{dp3dimdef}a. The gluing of the boundaries of the pieces of 
the original Riemann surface to form the daughter ones imply new adjacency 
relations between the unpaired pieces of the zig-zag paths. The result is 
two sets of consistent zig-zag paths, which can be used to construct 
the daughter dimer diagrams. This is shown in Figure \ref{dp3dimdef}b, 
where the two daughter dimer diagrams are seen to describe $N=4$ 
supersymmetric theories. This corresponds to two sets of D3-branes sitting 
at different smooth points in the deformed geometry, in agreement with the 
geometrical picture discussed above. 

This example makes manifest one subtle feature. In cutting the Riemann 
surface, it is not only the homology class of the 1-cycle that is 
important, rather it is crucial to use the particular representative 
associated with the tiling of $\Sigma$. In the above example the 
particular representative 
cuts the Riemann surface into two disks, while a more generic 
representative in the same homology class (e.g. that obtained by smoothing 
the intersections in figure \ref{dp3break}) would not yield that result.

This observation is important in a last respect. It is known that there 
exist several kinds of fractional branes at singularities, dubbed `$N=2$', 
`deformation' and `DSB' fractional branes in \cite{Franco:2005zu}. 
Different fractional branes lead to different behaviour of the infrared 
gauge theory. Interestingly, their differences become manifest in the 
brane tiling and in the associated 1-cycle in the mirror Riemann surface. 
Namely, only deformation branes lead to 1-cycles which separate the 
Riemann mirror surface into two pieces. Hence only deformation branes 
can lead to a splitting of the Riemann surface making the 1-cycle trivial, 
as required by the physical interpretation of the geometric transition.
For this connection to hold it is crucial {\em not} to allow to replace 
the 1-cycle determined by the tiling by another representative in the same 
homology class.

A last remark is in order. Notice that it is straightforward to derive the 
rule that complex deformations correspond to splitting off a sub-web in 
equilibrium. This follows from the fact that the cluster of dimer diagram 
faces associated to a deformation fractional brane have the topology of a 
disk. Hence, the total 1-cycle associated to the fractional brane has zero 
total $(p,q)$ charge. Thus the set of punctures it surrounds, and which 
correspond to the removed legs in the web diagram, have zero total 
$(p,q)$ charge, leading to a sub-web in equilibrium.

\subsection{Field theory interpretation}
\label{fieldcomp}

The gauge theory interpretation of complex deformations has been discussed 
for relatively general toric singularities in \cite{Franco:2005fd}, 
generalizing the discussion of the conifold in \cite{Klebanov:2000hb}.
Complex deformations of the geometry are related to confinement of the 
gauge factors associated to certain fractional branes. In 
\cite{Franco:2005fd}, the appearance of complex deformations in such 
situations was tested by introducing additional D3-brane probes. 
Confinement of the fractional brane gauge factors leads to a quantum 
deformed moduli space for these probes, hence reproducing the complex 
deformation of the underlying geometry. In fact, if the complex deformed 
space contains a left over singularity, the gauge theory on the D3-brane 
probe reduces, after confinement, to the quiver gauge theory associated to 
that singularity.

In this section we provide a description of the gauge theory analysis of 
the dynamics of the additional D3-brane probes, along the lines of 
\cite{Franco:2005fd}, in terms of dimer diagrams. We show that one can 
recover the dimer diagram of the left over singularities by simple 
operations in the dimer, with a clear gauge theory interpretation. 
Moreover, this recipe agrees with the procedure based on zig-zag paths 
described in Sections \ref{dconicomp}, \ref{moreexamplescomplex}.

Let us consider the example in Section \ref{dconicomp} of the double 
conifold deforming to a conifold. As discussed above, the complex 
deformation is triggered by introducing fractional branes changing the 
rank of the gauge factor 2. The complex deformation can be tested by 
introducing additional D3-branes probes. Hence we consider, as in 
\cite{Franco:2005fd}, the gauge theory with rank vector $(M,2M,M,M)$, and 
eventually we will focus on $M=1$.

\begin{figure}
\begin{center}
\epsfxsize=14cm
\hspace*{0in}\vspace*{.2in}
\epsffile{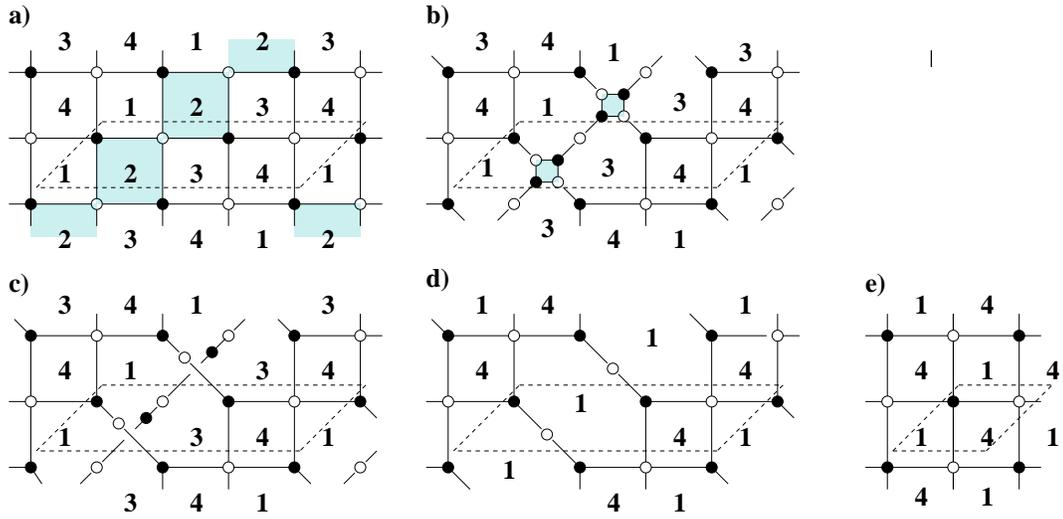}
\caption{\small 
Dimer diagram representation of the gauge theory analysis of D3-branes 
probing the complex deformation of the double conifold to the conifold.
(a) shows the fractional branes upon consideration. In (b) we introduce the 
mesons of the corresponding gauge factor. In (c) the gauge factor of the 
fractional branes confines and disappears. They leave behind a 
non-perturbative contribution to the superpotential of the mesons, 
implementing their quantum deformed constraint. The fact that the deformed 
space is not a toric variety is reflected by the fact that the dimer 
diagram is `non-planar' with edges passing through each other without 
intersection. In (d) we show the dimer diagram obtained when some mesons 
acquire a vev to saturate the quantum constraint. The resulting dimer 
diagram is equivalent to (e), which corresponds to the conifold theory.}
\label{dconifhu}
\end{center}
\end{figure}

For convenience we sketch the different steps in the gauge theory 
analysis (referring the reader to \cite{Franco:2005fd} for details), 
and their implementation in dimer diagrams, as shown in Figure 
\ref{dconifhu}. The starting dimer diagram with fractional branes is shown 
for convenience in Figure \ref{dconifhu}a.

The gauge factor 2 confines, hence one must introduce the corresponding 
mesons and baryons. The dynamics of the D3-brane probes is manifest on 
the mesonic branch, so the baryons will play no role and are set to zero. 
The mesons are however crucial so it is convenient to introduce them in 
an early stage. Figure \ref{dconifhu}b shows the gauge theory dimer 
diagram with the mesons of gauge factor 2 already introduced, even before 
taking into account the full effects of confinement. The mesons correspond 
to new edges appearing from the corners of face 2. This 
intermediate operation is formally similar to the implementation of 
Seiberg duality in dimer diagrams \cite{Franco:2005rj}.

Since the gauge factor 2 has the same number of colors and flavours, the 
classical constraint on the meson matrix has a quantum modification to 
${\rm det}\, {\cal M}=1$. This constraint can be implemented in the 
superpotential via a Lagrange multiplier. Actually, the D3-brane probe 
theory is manifest when the Lagrange multiplier has actually a non-zero 
vev. At this stage, all the non-abelian dynamics has been introduced and 
we can simply consider the abelian situation $M=1$. In this case, the 
non-perturbative contribution to the superpotential simply amounts to two 
quadratic terms in the mesons. The dimer diagram resulting from the 
above operations is shown in Figure \ref{dconifhu}c. The 
gauge factor 2 has confined and disappeared. It leaves behind the mesons 
with new interaction terms between mesons from opposite corners. The 
latter implies that the diagram cannot be drawn on a torus, since two 
edges must cross without intersection. This simply reflects the fact that 
the moduli space of the gauge theory is not a toric variety, in agreement 
with the fact that a complex deformation of a toric space is not itself a 
toric space. The dimer diagram clearly knows that after confinement a 
quantum deformation of the moduli space is taking place!

Although the full deformed geometry is not a toric space, one can zoom 
into a neighbourhood of the D3-brane probe and find a toric description 
for it. If one chooses mesonic vevs corresponding to locating the D3-brane 
probe at the left over singularity in the deformed space, one should then 
recover the gauge theory of D3-branes at this singularity. At the level of 
the dimer diagram, a choice of mesonic vevs saturating the constraint 
${\rm det}\, {\cal M}=1$ should lead to the dimer diagram of the left over 
singularity. In Figure \ref{dconifhu}d we show the result of this, 
where the removal of a line of edges corresponds to giving vevs to the 
associated mesons. After integrating out the bi-valent nodes, one obtains 
Figure \ref{dconifhu}e, which corresponds to the dimer diagram of the 
conifold theory. 

Hence, we have provided a set of simple dimer diagram rules which 
reproduce the gauge theory analysis of the D3-brane probe theory in 
\cite{Franco:2005fd}. Both techniques show that the geometry 
probed by the D3-brane is the complex deformation of the double conifold 
to the conifold.

Notice that the gauge theory operations we have just described nicely 
dovetail the procedure in terms of zig-zag paths described in previous 
sections. Namely the removal of the sub-web legs corresponds to 
confinement of the fractional brane. The edges 
which lose one of their zig-zag paths correspond to fields charged under 
the fractional brane group, and the process of closing them by new 
adjacency relations corresponds to constructing the mesons of the 
confining theory. More precisely, to the mesons that survive after 
satisfying the quantum constraint via introduction of vevs. Hence, the
zig-zag path method provides in one step the final dimer diagram of 
D3-branes at the left over singularity. Notice also the nice relation 
between our gauge theory process as described above, and the {\em ad hoc}
dimer diagram rules in \cite{Franco:2005fd}, see Figure \ref{dconiadhoc}.

\begin{figure}
\begin{center}
\epsfxsize=14cm
\hspace*{0in}\vspace*{.2in}
\epsffile{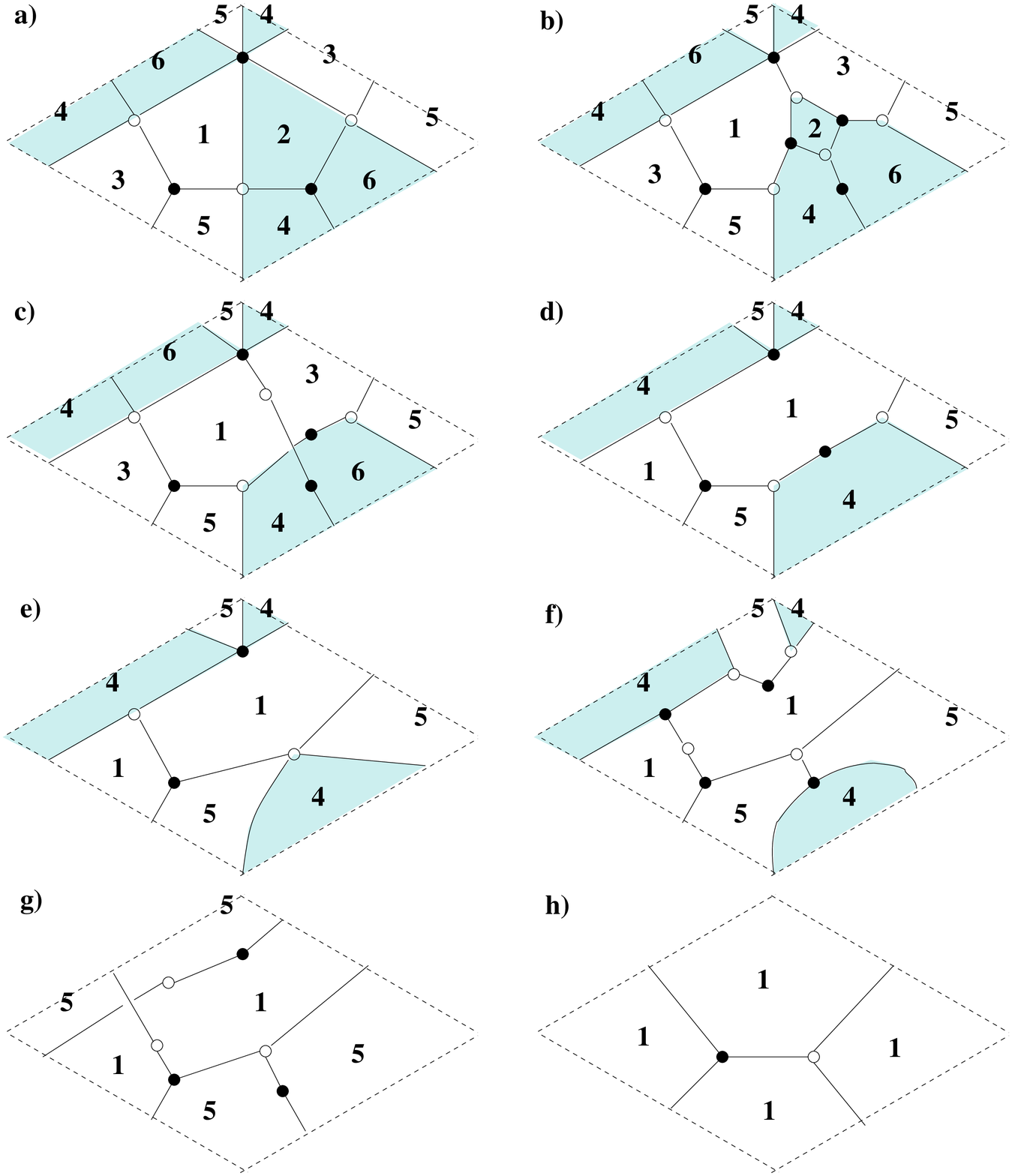}
\caption{\small 
Dimer diagram representation of the gauge theory analysis of D3-branes 
probing the complex deformation of the cone over $dP_3$ to a smooth 
geometry. (a) shows the fractional branes upon consideration. In (b) we 
introduce the mesons of the gauge factor 2. In (c) the gauge factor 2 
confines and induces a non-perturbative superpotential (associated to the 
quantum constraint on the mesons). In (d) the quantum constraint is 
saturated by giving vevs to certain mesons, which break the gauge factors 
1 and 3, and 4 and 6, to their diagonal combinations, respectively denoted 
1 and 4 in the following. After integrating out one bi-valent node (e), 
the gauge factor 4 has $N_f=N_c$, so we introduce its mesons (f). In (g) 
the gauge factor 4 confines and induces a non-perturbative superpotential. 
After giving vevs to some mesons, and integrating out bi-valent nodes, we 
obtain the dimer diagram (g), corresponding to D3-branes in a smooth 
geometry.}
\label{dp3fhu}
\end{center}
\end{figure}

We conclude this section by showing the dimer diagram representation of 
the gauge theory analysis of D3-branes probes in the complex deformation 
of the cone over $dP_3$ to a smooth space. The relevant steps are 
illustrated in Figure \ref{dp3fhu} and correspond to the gauge theory 
discussion in Section 4.3 in \cite{Franco:2005fd}. A sketch of this gauge 
theory analysis is provided along with the picture. The fact that the dimer 
diagram rules reproduce in a few easy steps an involved gauge theory 
analysis, such as this one, illustrates the power of these 
representations. Moreover, it is easy to show that the complete process is 
reproduced in a one-step fashion by the operations using zig-zag diagrams 
described in Section \ref{moreexamplescomplex}.

\subsection{Effect on perfect matchings}
\label{pmcomp}

In this section we describe complex deformations from another useful 
viewpoint, in terms of perfect matchings. As we show below, this has a 
very direct connection with the decomposition of the toric diagram as a 
Minkowski sum, used in the mathematical literature on complex deformations 
\cite{altmann1,altmann2,altmann3} (see also Appendix 
in \cite{Franco:2005zu} for a short description).

\begin{figure}
\begin{center}
\epsfxsize=14cm
\hspace*{0in}\vspace*{.2in}
\epsffile{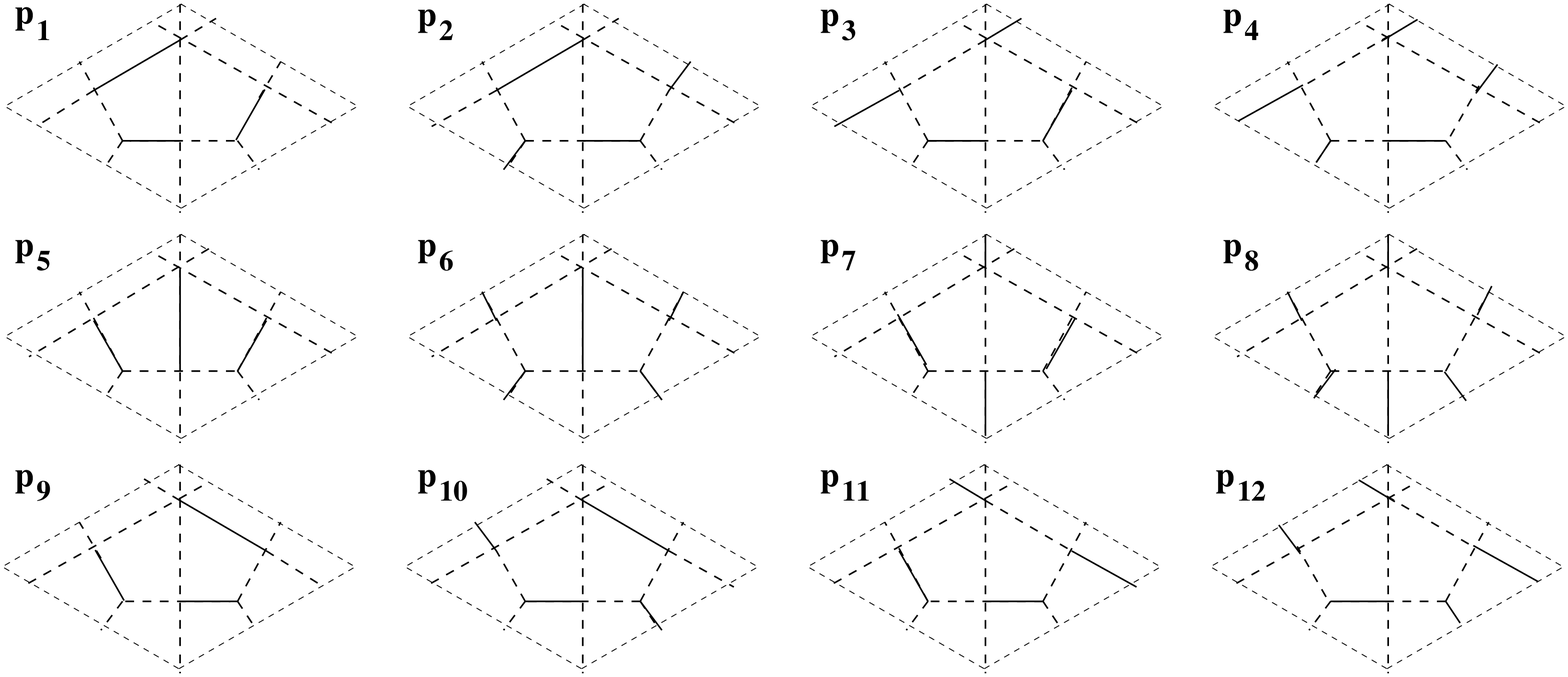}
\caption{\small Perfect matchings of the dimer diagram for the
$dP_3$ theory.}
\label{dp3pmatch}
\end{center}
\end{figure}

For concreteness let us consider an example which illustrates the general 
idea. Consider the deformation of the complex cone over $dP_3$ to a smooth 
space, as described in Section \ref{moreexamplescomplex}. The dimer 
diagram for the $dP_3$ theory is shown in Figure \ref{dp3dimer}, and 
its twelve perfect matchings are shown in Figure \ref{dp3pmatch}.

\begin{figure}
\begin{center}
\epsfxsize=4cm
\hspace*{0in}\vspace*{.2in}
\epsffile{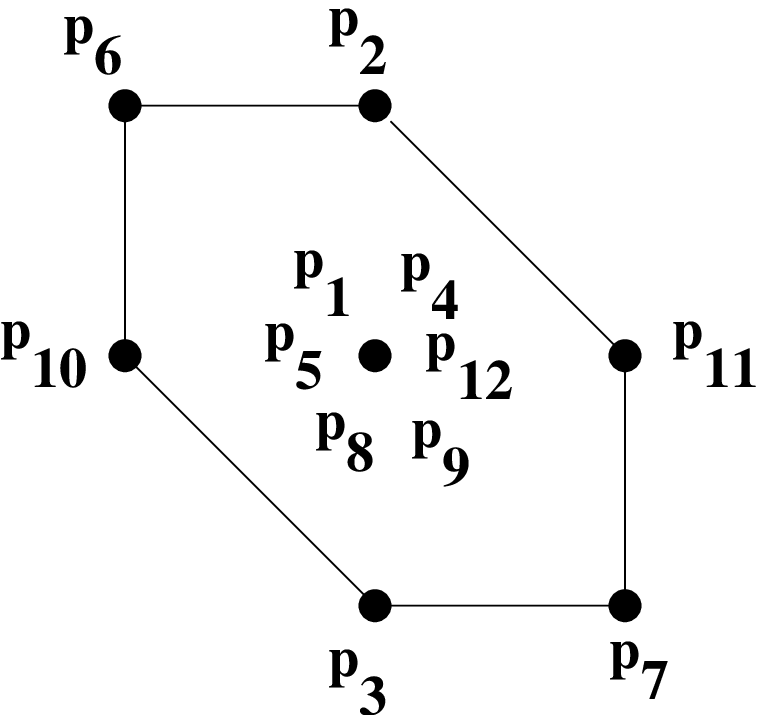}
\caption{\small Location of the perfect matchings on the toric diagram 
for the $dP_3$ theory.
}
\label{dp3heights}
\end{center}
\end{figure}

The location of these perfect matchings in the toric diagram, obtained 
as described in Section \ref{pmatchings} is shown in Figure 
\ref{dp3heights}a.

The effect of the complex deformation on the perfect matchings can be 
determined as follows. Recall that to each pair of perfect matchings we 
can associated a 1-cycle $p_j-p_i$ in $\Sigma$ obtained by superimposing 
them. The complex deformation amounts to cutting  $\Sigma$ in two pieces 
and gluing the boundary of each piece separately to obtain $\Sigma_1$ and 
$\Sigma_2$. In this process, some of the homology classes of $\Sigma$ 
become trivial in e.g. $\Sigma_1$. This implies non-trivial equivalence 
relations between difference paths of perfect matchings. For instance, 
matchings whose  difference path is a combination of the 1-cycles 
becoming trivial become equivalent in $\Sigma_1$. In addition to these 
identifications, the equivalence relations also imply changes in the 
slopes of the matchings, and thus an induced change in the toric 
diagram.

\begin{figure}
\begin{center}
\epsfxsize=8cm
\hspace*{0in}\vspace*{.2in}
\epsffile{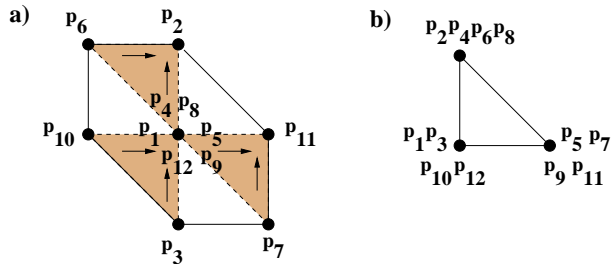}
\caption{\small 
a) The location of the equivalence classes of matchings describes the 
toric 
diagram of the first daughter geometry. In this case we recover the toric 
diagram of a smooth space, precisely the dual of the sub-web diagram 
corresponding to the legs B, D, F in Figure \ref{dp3complex}. Notice that 
all matchings in each vertex are equivalent (so their real multiplicity 
in the toric diagram is 1.
b) The equivalence relation between perfect matchings can be regarded as
the contraction of certain triangles in the toric diagram. 
} \label{dp3mink}
\end{center}
\end{figure}

To illustrate the procedure let us consider the deformation of the 
complex cone over $dP_3$ to a smooth space. In this example, studied 
above, $\Sigma_1$ is obtained 
by removing the zig-zag paths A, C, E, enclosed by the fractional brane 
associated to the gauge factors 2, 4, 6. It is easy to obtain the fate of 
the different 1-cycles of $\Sigma$ in $\Sigma_1$, by drawing them on the 
Riemann surface. For instance, the zig-zag paths A, C, E become trivial in 
$\Sigma_1$, whereas B, D, F remain non-trivial zig-zag paths in the 
daughter surface. Another interesting set of 1-cycles is given by those
associated to the $i^{th}$ gauge factor, which we denote by $f_i$. 
One can check that $f_2$, $f_4$, $f_6$ become trivial in $\Sigma_1$, while 
$f_1$, $f_3$, $f_5$ become identical to $-$B, $-$D, $-$F in $\Sigma_1$.
Then one can easily check relations like $p_1-p_3=-A-E+f_2\simeq 0$, etc. 
Following this, the matchings fall into three equivalence classes. We have
\beqa
p_1=p_3=p_{10}=p_{12} \quad  ; \quad p_2=p_4=p_6=p_8 \quad ; \quad
p_5=p_7=p_9=p_{11}
\eeqa
In addition one can obtain relations like $p_1-p_2=F+f_2\simeq F$, etc, 
which can be used to obtain the new relative slopes, and the positions of the 
(equivalence classes of) perfect matchings in the new toric diagram. The 
result of this operation is shown in Figure \ref{dp3mink}a. The result is
indeed the diagram corresponds to a smooth space, in fact the dual to the 
subweb corresponding to the legs B, D, F. The resulting diagram can be 
regarded as a contraction of the original toric diagram along a triangle, 
as illustrated in Figure \ref{dp3mink}b.

Clearly, one can operate in a similar manner with the other daughter 
Riemann surface $\Sigma_2$, and define another set of equivalences of 
perfect matchings. The end result is that the equivalence classes describe 
a toric diagram dual to the sub-web diagram corresponding to the legs A, 
C, E, as expected for $\Sigma_2$.

The splitting of the toric diagram into several sub-diagrams is dual to 
the splitting of the web diagram into sub-webs in equilibrium. Now
the splitting of the toric diagram into sub-diagrams, each of which
can be regarded as a contraction of the initial diagram (along the other 
sub-diagrams) has a nice connection with the mathematical operation known 
as decomposition of a toric polygon into Minkowski summands 
\cite{altmann1,altmann2,altmann3}. In fact, this operation lies at the 
heart of the mathematical characterization of complex deformations of 
toric singularities. It is a pleasant surprise to find a direct 
realization of the Minkowski decomposition of the toric diagram in the 
dimer language. We hope this relation brings the mathematics literature 
somewhat closer to the physical realization of deformations via geometric 
transitions (see \cite{Pinansky:2005ex} for an example of this).

As a final remark, we would like to point out that the use of perfect 
matchings also provides an interesting new way to recover the dimer 
diagrams of the daughter theories. Namely, after determining the 
equivalence classes of perfect matchings for a given Riemann surface, one 
can pick one representative of each class, and consider the set of edges 
involved in this set of matchings. The diagram obtained by superimposing 
all of them is exactly the dimer diagram that corresponds to the daughter 
theory.

\section{Conclusions}
\label{conclu}

In this paper we have provided tools to describe in full generality the 
processes that smooth out toric singularities, from the viewpoint of the 
gauge theory on D3-branes probing such geometries. The results are nicely 
cast in the language of dimer diagrams, in particular making use of the 
connection between the dimer diagram and the web diagram (via zig-zag 
paths and/or perfect matchings).

For partial resolutions our tools should allow a quick construction of the 
gauge theory for any toric singularity. It would suffice to implement out 
optimized tools using zig-zag paths to the partial resolution 
algorithm introduced in \cite{Morrison:1998cs}, with the advantage of 
being able to carry out a complicated partial resolution in one step. 

We have also provided a detailed gauge theory interpretation of the 
splitting of a dimer diagram into sub-dimers in the partial resolution 
process. It corresponds to a specific Higgs mechanism which splits the 
gauge theory into two gauge sectors decoupled at the level of massless 
states. We envision interesting model building applications of such 
systems.

An interesting open question is to understand the inverse process of 
combining different toric singularities into a single one, by inverting 
the partial resolution. Namely, by  adjoining the corresponding web 
diagrams along one or several legs, and to shrink the resulting segments. 
Progress in this direction should deal with ambiguities in the precise 
dimer diagrams to be combined, since the latter are defined modulo 
integration of bi-valent nodes.

For complex deformations we have provided a dictionary between the 
fractional branes and the precise splitting of the web diagram into 
sub-webs. Moreover, we have provided a simple set of dimer rules that 
reproduce the involved non-perturbative gauge theory analysis which 
describes D3-branes probes at such geometries. In addition the net result 
of these gauge theory operations is subsumed in extremely simple operations 
on the dimer in terms of zig-zag paths. It would be interesting to develop 
similar tools to analyze other infrared behaviours, like the removal of 
the supersymmetric vacuum by DSB branes.

We hope these tools are useful for these and other interesting purposes.

\centerline{\bf Acknowledgements}

We thank S. Franco and A. Hanany for illuminating conversations. A.M.U. 
thanks M.~Gonz\'alez for 
kind encouragement and  support. F.S. and I.G.-E. thank CERN TH for 
hospitality during completion of this project. This work has been 
partially supported by 
CICYT (Spain) under project FPA-2003-02877, and the RTN networks 
MRTN-CT-2004-503369 `The Quest for Unification: Theory confronts 
Experiment', and MRTN-CT-2004-005104 `Constituents, Fundamental Forces 
and Symmetries of the Universe'. The research of F.S. is supported by 
the Ministerio de Educaci\'on y Ciencia through an F.P.U grant. The 
research of I.G.-E. is supported by the Gobierno Vasco PhD fellowship 
program and the Marie Curie EST program.

\newpage

\appendix

\section{Proof of flatness}
\label{proof}

The flatness conditions can be checked in the general case, by a slight 
generalization of the analysis in the example in Section 
\ref{fieldtheory}. We recall that in this section we are considering
original dimer diagrams not containing bi-valent nodes (hence they have 
been integrated out if originally present).

{\bf F-flatness conditions:} 
As described in Section \ref{matchings}, in partial resolutions each 
subdimer contains at least one perfect matching of the original dimer 
diagram. This implies that every sub-dimer contains all the nodes of the 
original dimer diagram. From this follows that in any sub-dimer, for any 
node there are at least two edges ending on it in every sub-dimer. At 
the level of the gauge theory, this implies that for each superpotential 
term of the original theory there are a sufficient number of  
bi-fundamentals with zero vevs to automatically satisfy the F-term 
conditions. Hence the assignment of vevs dictated by the dimer rules is 
F-flat.

{\bf Non-abelian D-flatness conditions:}
As described in section \ref{splitting}, we divide the
set of zig-zag paths into two disjoint sets, where each set admits a
dual interpretation as the set of external legs in the web diagram
that we take to infinity. Let us denote collectively the elements
belonging to the first set as 1 and those belonging to the other set
as 2.

\begin{figure}
\begin{center}
\psfrag{pl1}{1}
\psfrag{pl2}{1}
\psfrag{label}{(1)}
\includegraphics{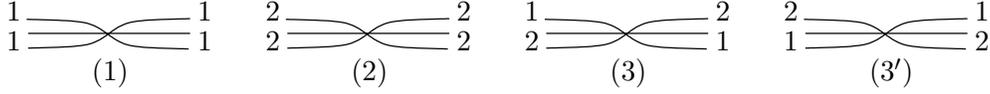}
\psfrag{pl1}{2}
\psfrag{pl2}{2}
\psfrag{label}{(2)}
\includegraphics{zzedge.eps}
\psfrag{pl1}{1}
\psfrag{pl2}{2}
\psfrag{label}{(3)}
\includegraphics{zzedge.eps}
\psfrag{pl1}{2}
\psfrag{pl2}{1}
\psfrag{label}{($3'$)}
\includegraphics{zzedge.eps}

\caption{\small The four possible types of edges, classified according to the
zig-zag paths meeting at the edge.}
\label{zzedge}
\end{center}
\end{figure}

Consider a given face in the dimer diagram, and orient its edges by 
running through them e.g. counterclockwise. Each edge can then be 
classified into 4 types depending on which kind of zig-zag paths intersect 
over it. We will denote the four kinds as type 1, 2, 3 and $3'$, see 
figure \ref{zzedge}, where 3 and $3'$ are distinguished by the 
orientation \footnote{The similar notation for edges and zig-zag paths is 
introduced to (hopefully) improve the readability. In the rest of this 
section we mostly deal with edges, so this should not cause too much 
confusion.}.

\begin{figure}
\begin{center}
\psfrag{pl1}{2}
\psfrag{pl2}{1}
\psfrag{pl3}{2}
\psfrag{pl4}{2}
\includegraphics{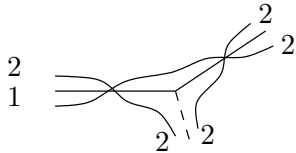}

\caption{\small Inconsistent pasting of edges ($3'2$). Note that the only
  constraints come from the joining of the interior zig-zag paths. The
  exterior ones can be arbitrary as they do not need to be joined (in
  the absence of bi-valent nodes) since they ``run off'' along some extra
  edge, denoted by the dashed line in the drawing.}
\label{zzbadedge}
\end{center}
\end{figure}

In this fashion, we assign to each face a (periodic) string
of symbols given by the kind of edges we encounter when traversing the
face counterclockwise. A typical string will then look like:\[
\ldots 3'1323'3 \ldots \] where we have written just the period. It is 
easy to realize that any valid string should satisfy a few rules which we 
can read from the dimer diagram. Namely there are some sequences of 
symbols that are not allowed, for example $3'2$. To see this, focus on the 
zig-zag paths ``interior'' to the edge. The given sequence would tell us 
that a type 1 zig-zag path exits the $3'$ vertex from the right, and then 
joins a type 2 zig-zag path in the next edge, see figure \ref{zzbadedge}. 
This is obviously not allowed. The other disallowed sequences are
$13'$, $23$, $31$, $33$, $3'3'$, $12$ and $21$.

We can then associate to the most general face in a dimer a sequence of 
symbols not containing these forbidden words. It is easy to convince 
oneself that in any such string, at least one of the following 
substitutions applies and gives rise to another consistent sequence with 
two symbols removed in the period (``$\cdot$'' denotes the empty word):
\begin{eqnarray*}
11 & \longrightarrow  \cdot  \quad\quad ; \quad\quad & 22  \longrightarrow  
\cdot 
\quad\quad ; \quad\quad 33' \longrightarrow  \cdot \quad\quad ; 
\quad\quad  3'3  \longrightarrow  \cdot \\
132 & \longrightarrow  3 \quad\quad ; \quad\quad
& 23'1  \longrightarrow  3'
\end{eqnarray*}
As an example, applying the rules one would get the following sequence
of strings:\[3'133'1132 \lra 3'11132 \lra 3'132 \lra 3'3 \lra \cdot\]

Since we can always apply one of these rules, and all of them reduce
the length of the string by two, we have found that it is always 
possible to reduce an arbitrary string to nothing \footnote{The sequences 
always have even length, consistently with anomaly cancellation.}.
The interesting fact about these operations is that on the field theory 
side they do not change the value of the D-term. Essentially, the first 
four rules preserve the D-term value because the disappeared edges 
correspond to a fundamental and an antifundamental with the same vev, 
hence with canceling contributions to the D-term. For the last two rules,
the disappeared edges have vevs whose contributions add up to the trace of 
an $SU(N)$ generator, which is zero. One can in this way easily translate 
between the language used in equation \ref{dflat} and this language of 
sequences. What this means is that the value of the D-term for all 
possible faces in a dimer is given by the D-term of the empty sequence, 
which is equal to zero.

\begin{figure}
\begin{center}
\psfrag{pl1}{2}
\psfrag{pl2}{1}
\psfrag{pl3}{2}
\psfrag{pl4}{2}
\psfrag{pl5}{1}
\psfrag{pl6}{2}
\includegraphics[scale=.5]{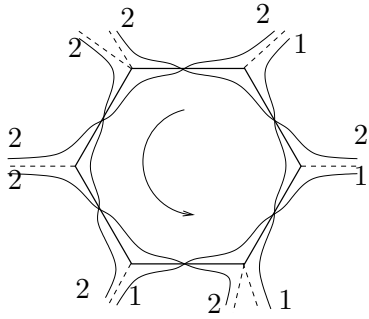}

\caption{\small A possible face in a dimer, where we have indicated the
  relevant classification for the zig-zag paths. External edges are
  denoted by the dashed lines, and the arrow indicates the traversal
  direction used in the text when enumerating the edges.}
\label{dtermface}
\end{center}
\end{figure}

As an example, let us study the configuration depicted in figure
\ref{dtermface}. The periodic string we associate with the face is
given by $\ldots 223'132 \ldots$. Applying the rules we have described
a possible reduction to nothing would be:\[
223'132 \lra 23'32 \lra 22 \lra \cdot
\]
This proves that the D-term for the relevant gauge group
vanishes.

\end{document}